\documentclass[onecolumn,showpacs,preprintnumbers,amsmath,amssymb]{revtex4}

\usepackage{epsfig}

\newcommand{\D}{\mathrm{d}}

\newcommand{\E}{\mbox{1\hspace{-3 pt}I}}

\newcommand{\Exp}[1]{\text{e}^{#1}}

\newcommand{\pd}{\partial}

\newcommand{\Hr}{\bigcirc \hspace{-7.5 pt} r \hspace{2 pt}}
\newcommand{\Hc}{\bigcirc \hspace{-7.5 pt} c \hspace{2 pt}}
\newcommand{\hr}{\bigcirc \hspace{-6 pt} r \hspace{2 pt}}
\newcommand{\hc}{\bigcirc \hspace{-6 pt} c \hspace{2 pt}}
\newcommand{\gopr}{g^{\text{u}}_{\text{c} \hr \rightarrow  }}
\newcommand{\gonr}{g^{\text{u}}_{\text{c} \hr \leftarrow  }}
\newcommand{\gbpr}{g^{\text{b}}_{\text{c} \rightarrow \hr  }}
\newcommand{\gupr}{g^{\text{u}}_{\text{c} \rightarrow \hr  }}
\newcommand{\gbnr}{g^{\text{b}}_{\text{c} \leftarrow  \hr  }}
\newcommand{\gunr}{g^{\text{u}}_{\text{c} \leftarrow  \hr  }}
\newcommand{\gbpc}{g^{\text{b}}_{\hc \rightarrow \text{r} }}
\newcommand{\gupc}{g^{\text{u}}_{\hc \rightarrow \text{r} }}
\newcommand{\gbnc}{g^{\text{b}}_{\hc \leftarrow  \text{r} }}
\newcommand{\gunc}{g^{\text{u}}_{\hc \leftarrow  \text{r} }}
\newcommand{\gbr}{g^{\text{b}}_{\hr}}
\newcommand{\gur}{g^{\text{u}}_{\hr}}
\newcommand{\gbc}{g^{\text{b}}_{\hc}}
\newcommand{\guc}{g^{\text{u}}_{\hc}}
\newcommand{\gcwet}{g^{\text{wet}}_{\hc}}
\newcommand{\gc}{g^{\text{at}}_{\text{c}}}
\newcommand{\gcwall}{g^{\text{at}}_{\text{wall}}}
\newcommand{\gcHSS}{g^{\text{at}}_{\hc}}
\newcommand{\bs}{\gamma^{\text{b}}}
\newcommand{\us}{\gamma^{\text{u}}}
\newcommand{\ubs}{\gamma^{\text{u/b}}}
\newcommand{\sA}{s_{\text{A}}}
\newcommand{\sB}{s_{\text{B}}}
\newcommand{\gA}{g_{\text{A}}}
\newcommand{\gB}{g_{\text{B}}}
\newcommand{\scrit}{s_{\text{crit}}}
\newcommand{\gvol}{\gamma_{\text{vol}}}
\newcommand{\Ecb}{E_{\text{cb}}}
\begin{document}

\title{Equation of State of Wet Granular Matter}

\author{A. Fingerle}
\email{axel.fingerle@ds.mpg.de}
\author{S. Herminghaus}
\email{stephan.herminghaus@ds.mpg.de}
\affiliation{Max-Planck-Institute for Dynamics and
Self-Organization, Bunsenstr. 10, 37073 G\"ottingen, Germany}

\date{\today}


\begin{abstract}
A novel expression for the near-contact pair correlation function
of $D$-dimensional hard sphere systems is presented which arises
from elementary free-volume arguments. Its derivative at contact
agrees very well with our simulations for $D=2$. For jammed
states, the expression predicts that the number of exact contacts
is equal to $2D$, in agreement with established simulations. When
the particles are wetted, they interact by the formation and
rupture of liquid capillary bridges. Since formation and rupture
events of capillary bonds are well separated in configuration
space, the interaction is \emph{hysteretic} with a characteristic
energy loss $\Ecb$. The pair correlation is strongly affected by
this capillary interaction depending on the liquid-bond status of
neighboring particles. A theory is derived for the nonequilibrium
probability currents of the capillary interaction which determines
the pair correlation function near contact. This finally yields an
analytic expression for the equation of state, $P=P(N/V,T)$, of
wet granular matter for $D=2$, valid in the complete density range
from gas to jamming. Driven wet granular matter exhibits a
van-der-Waals-like unstable branch at granular temperatures
$T<T_{\text{c}}$ corresponding to a first order segregation
transition of clusters. For the realistic rupture length of the
liquid bridge, $\scrit=0.07 d$, the critical point is located at
$T_{\text{c}} = 0.274 \Ecb$. While the critical temperature weakly
depends on the rupture length, the critical density
$\phi_{\text{c}}$ is shown to scale with $\scrit$ according to
$\scrit = 4d (\sqrt{\phi_{\text{J}}/\phi_{\text{c}}}-1)$. The
segregation transition is closely related to the precipitation of
granular droplets reported for the free cooling of one-dimensional
wet granular matter \cite{Fingerle2006}, and extends the effect to
higher dimensional systems. Since the limiting case of sticky
bonds, $\Ecb \gg T$, is of relevance for aggregation in general,
simulations have been performed which show very good agreement
with the theoretically predicted coordination $K$ of capillary
bonds as a function of the bond length $\scrit$. This result
implies that particles that stick at the surface, $\scrit=0$, form
isostatic clusters. An extension of the theory in which the bridge
coordination number $K$ plays the role of a self-consistent
mean-field is proposed.
\end{abstract}

\maketitle

\section{Introduction}

Dry sand trickles easily through chinks and crevices, as everyone
knows well from the hour glass, or just personal experience.
However, the addition of small amounts of liquid are sufficient to
transform it into a plastic (or, more precisely, a viscoplastic)
material. The same is true for all granular matter when a few
volume percent of liquid are added, provided the latter wets the
grains well and the grains are not too large. It is understood
that this dramatic change, from a quasi-fluid to a solid behavior,
is due to the formation of liquid bridges
\cite{Rum62,Thor91,ThorYin91,Lian93,Sim93,Sim94,Thor96,Lian98,Wil00,Herm05}
between the granules wherever they come into contact. These liquid
bridges mediate a cohesion force, and rupture as soon as the
particle surfaces are separated by a distance $\scrit$ which
scales as the cube root of the amount of added liquid
\cite{Herm05}. These processes of formation and rupture of liquid
bridges are the main cause of the observed dramatic changes in the
mechanical properties of the material. Because of the generality
of the effects, it has become common to study systems with
spherical grains (usually glass beads), in order to ease
theoretical modelling and to avoid  side effects. We decided to
follow this approach.

In this work we show analytically that the peculiar interaction by
capillary bridges gives rise to a first order transition, and we
compute the critical density and the critical temperature. We
shall focus on the two dimensional case, but many concepts carry
over to dimensionality $D=3$. Since there is no clear observation
of a first order phase transition in the hard-sphere fluid for $D
\leq 2$ \cite{Binder2002,Mak2006}, the added liquid leads to a
qualitative change. More importantly, this transition is
determined entirely by the geometric and energetic properties of
the capillary bridges.

A dry system of $N$ hard spheres with diameter $d$ confined to an
area or volume $V$ has no intrinsic energy scale, so that the
equation of state is of the form $P=T \ f(N/V)$ with the
temperature $T=\left<m v_i v_i\right>$ and a nonlinear density
dependence, $f$. The defined size of hard particles is
conveniently used to restate the density $n=N/V$ as the
dimensionless occupied fraction $\phi=\sigma_D {n d^D}/({2^D} \
D)$ ($\sigma_D$ the surface of a $D$-dimensional unit sphere),
which is the area fraction $\phi=\frac{\pi}{4} n d^2$ for two,
and volume fraction $\phi=\frac{\pi}{6} nd^3$ for three
dimensions.

\begin{figure} \begin{center}
(A) \epsffile{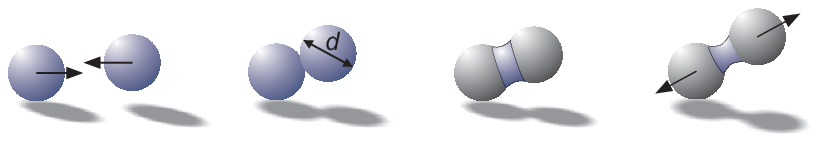} \\ (B)
\scalebox{0.71}{\epsffile{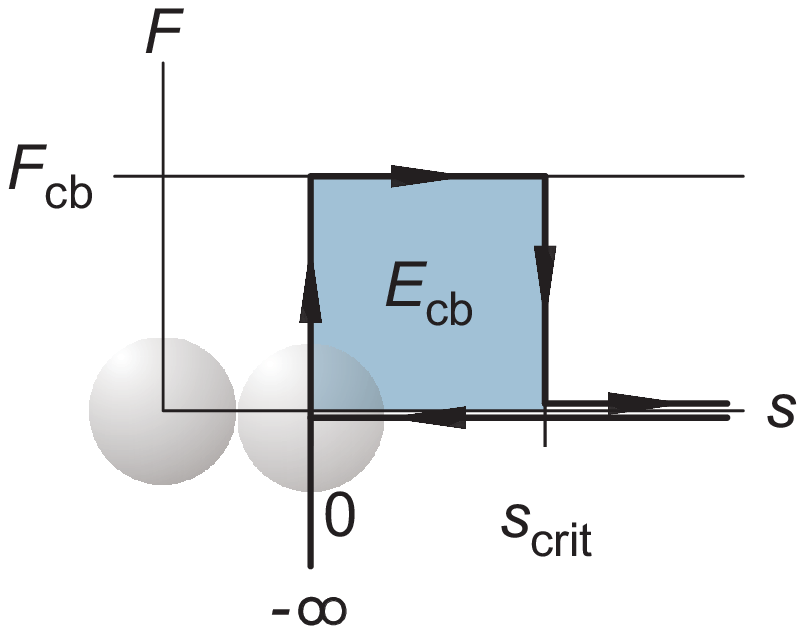}} \caption{The
hysteretic interaction in wet granular matter. (A) Capillary
bridges form at contact and mediate an attractive force
$F_{\text{cb}}$. At the bridge length $\scrit$ the bridge becomes
unstable and pinches off. (B) This hysteretic interaction by
capillary bridges gives rise to a well-defined loss of energy
denoted by $\Ecb$. The rupture length $\scrit$ is largely
exaggerated for illustration. While the particle diameter $d$ is
the only length scale for dry granulates, in wet granular matter
there is a second scale set by $\scrit$. A realistic value
is $\scrit \approx 0.07 d$, which is realized when $1\%$ of the
jamming volume is added by a wetting liquid (with zero contact
angle). Furthermore, the bond energy $\Ecb$ defines an intrinsic
energy scale, which is absent in dry granulates. As it is shown,
the length and energy scale set by the capillary interaction give rise to a
phase transition with a critical density $\phi_{\text{c}}$
and a critical granular temperature $T_{\text{c}}$.}
\label{chpEqStGraphic1}
\end{center} \end{figure}

The capillary interaction of wet granular matter has a
well-defined binding energy $\Ecb$ \cite{Wil00}, and it has been
demonstrated experimentally \cite{FingerleEcb} under realistic
dynamical conditions with impact velocities typical for strongly
fluidized wet granular matter that the hysteretic character of the
interaction is essential: the dominant mechanism of dissipation is
the hysteretic formation and rupture of capillary bridges, the
energy $\Ecb$ of which is irreversibly taken from the kinetic
energy of the granular motion whenever a liquid bridge ruptures
\cite{Herm05}. The bridge energy has been quantified
\cite{Herm05,FingerleEcb}, according to which $\Ecb$ is
proportional $d \sqrt{W}$. Figure~\ref{chpEqStGraphic1}
illustrates this hysteresis in the Minimal Capillary model
\cite{Herm05} applied here, which assumes a constant bridge force
$F_{\text{cb}}$. This may appear as an oversimplification at first
glance, but there is increasing experimental evidence that the
details of the force law are insignificant for the collective
dynamics on which we focus here
\cite{Herm05,FingerleCoEx,FingerleEcb}, as also confirmed from the
point of view of dynamical systems theory \cite{Fingerle2005,
Fingerle2007}.

Obviously, an external energy current has to be continuously
injected to drive the system into a nonequilibrium steady state.
In the equilibrium limit, $\Ecb \rightarrow 0$, we will have a
pressure of the form $P=T \ f(\phi)$. It is the objective of this
article to derive the equation of state for the hysteretic liquid
bridge interaction of wet granular matter in such a driven state.
In view of the intrinsic energy scale $\Ecb$, this relation has to
be of the form $P = P(\phi,T/\Ecb)$.

The equation of state is understood as an intrinsic property of
homogeneous wet granular matter, kept in a stationary
nonequilibrium state of granular temperature $T$. With this given
temperature we may subsume various ways in which the system can be
externally driven to compensate for the dissipation by rupturing
liquid bridges, so that this granular temperature $T$ is
maintained over many particle diameters.

We remark that in most experimental situations involving wet
granular matter, the granular temperature is a nonlinear, even
discontinuous, response depending on the details of the driving,
such as boundary motion or air flow in air-fluidized beds. In this
article we deliberately regard the granular temperature as the
\emph{control parameter}, so that the theoretical description of
the boundary coupling is conveniently separated. Yet we emphasize
that for the full description of an experimental situation one has
to insert the equation of state into the equation for the external
energy input, and then solve for the granular temperature as the
\emph{nonlinear response} to the external driving.


We aim at describing the steady nonequilibrium states of wet
granular matter, which are so multifaceted that at first glance
one might think that aside from density and granular temperature
further physical parameters are necessary in order to describe
such a state. Yet as simulations have shown, states of wet
granular matter far from equilibrium \cite{Fingerle2006} are very
well described by a single granular temperature $T$ assuming a
Gaussian velocity distribution, neglecting higher cumulants.
Furthermore, it is known that the self-organized velocity
distribution of free cooling wet granular matter has a vanishing
fourth cumulant \cite{Vasily2006}. We point out that the condition
of a locally isotropic and homogenous state used in this work
implies that the temperature field may vary only slowly over many
particle diameters so that there is no strong influence by a heat
current, which would otherwise be considered as a third parameter
of the local nonequilibrium state.

Throughout this study, we allow for a certain polydispersity, $0
\leq \Delta d / d < 0.1$. (For higher polydispersity, the dense
system undergoes a kinetic glass transition
\cite{SantenKrauth2000, SantenKrauthOnline}). First of all,
polydispersity is frequently used in simulations and experiments
to prevent the monocrystalline state. Secondly, most systems of
practical relevance exhibit some polydispersity. Another
characteristic of 'real' granulates is that the surfaces of the
grains are not ideal, bearing certain roughness. This does,
however, only change the amount of liquid which must be added in
order to achieve the capillary interaction: first some liquid is
required to fill the crevices and tiny recesses in the grain
surfaces, until the grains effectively have a smooth liquid
coating, which is then completely wetted by all additional liquid.
For glass beads, as those used in most of the experiments, this is
typically the case above a volume fraction $W_{min} = 0.1 \%$ of
liquid with respect to the jammed granular sample volume. We also
require an upper limit on the volume fraction of the wetting
liquid, so that the maximal length $\scrit$ of liquid bridges is
of the order or below the polydispersity $\Delta d$ of the
spheres. This is to demand that $\scrit/d \approx \sqrt[3]{W}/3$
is smaller than $\Delta d / d < 0.1$, so that
$W<W_{\text{max}}=2.8 \%$. This happens to closely coincide with
the upper limit set on the liquid content to ensure that
neighboring capillary bridge do not merge
\cite{MarioNatureMaterials}. For this range of the liquid content
the capillary interaction is a truly pairwise interaction with the
capillary force acting radially between pairs of particles.
Another implication of roughness is that there is a substantial
tangential friction between adjacent grains. This means that in
principle one has to include all rotational degrees of freedom in
the kinetic considerations for any statistical physical treatment
of our system. However, we are here focussing on the effects due
to the liquid capillary bridges, which mediate central forces.
These do not couple to the (tangential) rotational modes. We
therefore expect that the rotational degrees of freedom play, in
our system, the role of a spectator heat bath which follows the
translational dynamics, but does not influence it greatly, aside
from a quantitative increase of the granular specific heat. In
fact, recent experiments and simulations of wet granular systems
\cite{FingerleCoEx} show that this approach yields remarkable
agreement with experimental data. In this work, we thus completely
neglect all rotational degrees of freedom.

\section{Dry Spheres as Starting Point}
Before we add the wetting liquid to the hard sphere system, we
investigate the dry case in this section and derive expressions
for the pair correlation near contact, which will be extended to
the wet case in the following section.

\begin{figure} \begin{center}
\epsffile{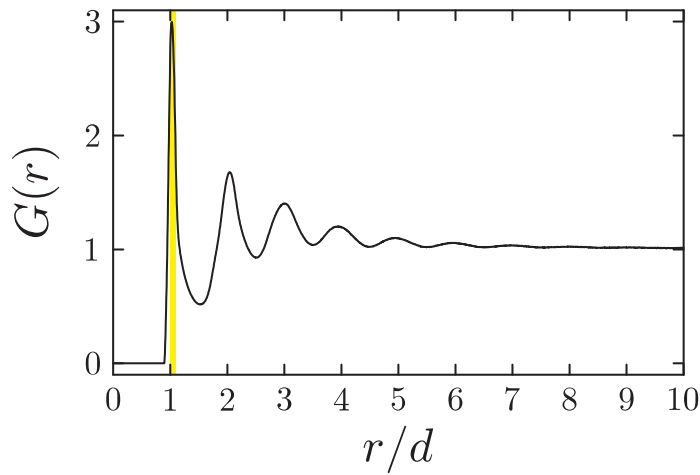} \caption{The pair correlation of
wet granular matter in a fluidized state resulting from a
molecular dynamics-type simulation in $D=2$ dimensions. The
correlation function $G(r)$ vanishes in the range $(0,{d})$ where
the finite particle size leads to excluded volume. We use the
function $g(s)$ with the surface separation $s$ of neighboring
particles as it is convenient for wet granular matter where
interstitial liquid bridges have the length $s$. Note that this is
not exactly identical to the function $G(d+s)$ shifted by one
particle diameter $d$, since a realistic granular system has some
polydispersity $\Delta d$ around the mean diameter ${d}$. Aside
from kinetic contributions, the pressure is due to the interaction
forces which become dominant with increasing density. The internal
forces in wet granular matter are short-ranged. Therefore our
interest focuses on the sharp fall-off in the indicated range
$0<s<\scrit$ of capillary interaction. This highlighted region
indicates the typical range of $\scrit$, and corresponds to the
region highlighted in Fig.~\ref{chpEqStgAgB}. Furthermore, we
derive more detailed correlation functions, $g^{\text{u}}(s)$ and
$g^{\text{b}}(s)$, for unbound and capillary connected pairs,
respectively, in order to describe the hysteretic interaction in
wet granular matter.} \label{chpEqStGraphic2}
\end{center} \end{figure}

Due to their finite size, the positions of hard spheres are not
distributed independently from each other, as it is the case for
the point-particles of the ideal gas. The configuration space of
$N$ spheres is not $V^N$, but restricted to a concave subset in
which the systems moves chaotically as a high dimensional
billiard. With the absence of an intrinsic energy scale, the dry
system is athermal, which means that a change in temperature is
equivalent to rescaling the time axis. The excluded volume gives
rise to correlations in the particle positions, which are measured
by the pair correlation function. Denoting by $n=N/V$ the mean
macroscopic particle density and by $n_{\text{m}}({\bf
r})=\sum^{N}_i \delta({\bf r}-{\bf r}_i)$ the microscopic density,
the isotropic pair correlation $G(r)$ is defined as the
probability
\begin{eqnarray}
\left< n_{\text{m}} ({\bf r}) \right>_{\text{particle at }0} \
\D\,\text{vol} = n \ G(\vert {\bf r} \vert) \ \D\,\text{vol}
= n \ g(s) \ \D\,\text{vol}
\label{DefPairCorr}
\end{eqnarray}
to find the center of a particle in the shell
$\D\,\text{vol}=\sigma_D \; r^{{(D-1)}} \; \D r$ of radius
$r=r_i+r_j+s$ and thickness $\D r = \D s$ centered around a
reference particle. We have conveniently subtracted the particle
radii $r_i+r_j$ in the last equality of (\ref{DefPairCorr}), so
that $s>0$ is the surface separation.
The function $g(s)$ is advantageous
for polydispersity scattered around the mean diameter $d$
\footnote{We do not distinguish between the mean value and the
root mean square of the diameter (relevant to the Vorono\"i area),
because they differ only by $(\Delta d/d)^2/2<5\times10^{-3}$.},
because of its defined contact point, $s=0$, which is
smeared out in the function $G(r)$. Furthermore it is the
natural way to describe an interstitial liquid bridge between the
considered pair of particles, with $s$ the length of the bridge.
For a certain liquid volume per particle and contact angle of the
wetting liquid, there is a well defined critical bridge length
$\scrit$ at which the bridge becomes unstable and
ruptures. The mean density $n$ is factored out in
(\ref{DefPairCorr}) so that the dimensionless $g$ would
be equal to unity for all separations if there was no particle-particle
correlation. Figure~\ref{chpEqStGraphic2} shows the pair
correlation of a fluidized
state in which long range order is lost,
so that $g(s)$, respectively $G(r)$, tends to
unity for $r \gg d$.

The forces in wet granular matter, hard-core repulsion and liquid
bridge attraction, are short-ranged and radial, acting between
pairs of particles over a separation range $0<s<\scrit$ with
$\scrit \ll d$. We are therefore interested in the short-range
behavior of the pair correlation $g(s)$ up to leading order in
$s/d$. For such short particle separations the pair correlation
$g(s)$ is (up to a normalization constant) just the probability to
find next neighbors at a separation $s$. Put in equivalent words:
decomposing the pair correlation function $g(s)=\sum_{k=1}^\infty
g_k(s)$ in contributions $g_k$ of the $k$'s shell of Vorono\"i
neighbors, we have $g(s)=g_1(s)$ in the range of interest,
$0<s<\scrit \ll d$. To shorten notation we suppress the subindex
1.

\begin{figure} \begin{center}
\epsffile{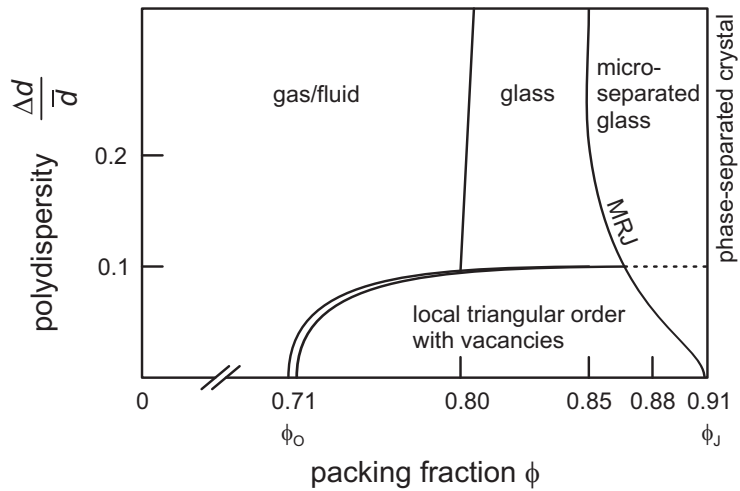} \caption{This plot reviews
\cite{SantenKrauthOnline}, \cite{Donev2007} (Fig.~15 therein), and
\cite{TakanaPrivCom}. The athermal transitions shown are
properties of the configuration space of hard discs. Since the wet
granular dynamics takes place in this configuration space, results
for hard discs form the starting point for a theory of wet
granular matter in two dimensions. At low (gas) and moderate
(fluid) densities $\phi$, the configuration space is probed
ergodically and the system has low shear viscosity. As density is
increased, the system gets trapped in a disordered state (glass
for polydispersity above 0.1), or in a state with local triangular
order. Both transitions, the glass transition (vertical line at
$\phi\approx 0.80$) and the ordering transition (curved line
ending at $\phi_{\text{o}}=0.71$) can be detected by the rapid
increase of the shear viscosity $\eta$ (cf.~\cite{LudingTransport}
for the ordering transition). While there is an athermal first
order transition in three dimensions, it is at present discussed
in the literature whether the transition region (double lines
ending at $\phi_{\text{o}}$) represents a fluid/solid coexistence
(corresponding to a weak first order transition with a small jump
of the entropy per particle) or if there is an intermediate
hexatic phase (according to the
Kosterlitz-Thouless-Halperin-Nelson-Young scenario)
\cite{Binder2002,Mak2006}. As the packing fraction $\phi$ is
increased further, the islands to which the system is confined in
the configuration space shrink to points. This jamming limit can
be detected by the divergence of the pressure $p$ (at fixed
granular temperature) under compression (for example using the
particle expansion of the Lubachevsky-Stillinger algorithm). The
maximal random jammed state (MRJ, for strict jamming as defined by
Torquato, Truskett, and Debenedetti~\cite{Torquato2000}) is the
vertical curve at the right. Densities higher than MRJ are deep in
the glassy regime. From the thermodynamic point of view, the
system would cluster in phases separated according to the particle
size, but this eutectic freezing-transition is kinetically
suppressed \cite{Donev2007} and unreachable.}
\label{chpEqStGraphic2b}
\end{center} \end{figure}

\subsection{The Dense Limit} \label{DryDense}
Figure~\ref{chpEqStGraphic2b} gives an
overview of results by
\cite{SantenKrauthOnline},
\cite{Donev2007} (Fig.~15 therein),
and \cite{TakanaPrivCom}
for the phases of the two-dimensional system
depending on density and polydispersity.
For polydispersity below 0.1, there are two density regimes
separated by the ordering transition at 
$\phi_{\text{o}}$ \footnote{Numerics reported at
present \cite{Binder2002,Mak2006}
could not distinguish between a weak first order transition (with
constant pressure at $\phi_{\text{o}}$), and two subsequent
continuous transitions (with slowly increasing pressure at
$\phi_{\text{o}}$).}.
These transitions are
a purely geometric property (i.e. excluded volume effect)
of the configuration space and are therefore
athermal. To compute the radial next-neighbor distribution
at densities above the critical density,
$\phi>\phi_{\text{o}}$, we consider the Vorono\"i tessellation of
the system, which embeds each particle into a convex polygonal
cell.
The sizes $\{V_i\}$ of these Vorono\"i cells scales as $(d+s)^D$, with the
particle separation $s$. The mean cell size $\sum_i^N V_i/N = V /
N = 1/n$ is exactly the inverse density $n$. Hence,
\begin{eqnarray}
\left< \left(1+\frac{s}{d} \right)^D \right>  =
\frac{n(s\rightarrow 0)}{n} =
\frac{\phi(s\rightarrow 0)}{\phi} \ , \ \label{DensityCondition}
\end{eqnarray}
where the triangle brackets denote averaging over next neighbors
which are in contact ($s \rightarrow 0$) with the center particle
at the jamming density $\phi(s \rightarrow 0)$.
We refer to those pairs of particles which come into contact
at jamming as neighbors of type A,
i.e the surface separation $\sA$ of A-neighbors is
\begin{eqnarray}
\sA = 0 \text{ at } \phi=\phi_{\text{J}} \ . \label{JammingOfA}
\end{eqnarray}
In the monodisperse limit for $D=2$, $\phi_{\text{J}}$ assumes the value of
the triangular crystal, $\phi_{\text{max}}=\pi/(2\sqrt{3})=0.91$.
Polydispersity decreases the (maximal random) jamming density
$\phi_{\text{J}}$ and
increases the critical density $\phi_{\text{o}}$ for the onset of
triangular order as shown in
Fig.~\ref{chpEqStGraphic2b}.




\subsubsection{Contribution to the Contact Correlation: The A-Neighbors}

Since the Vorono\"i cells exchange their free volume, $V-V_{\text{min}} \propto
\left(1+\frac{s}{d} \right)^D - 1$, and the total volume is
conserved we assume an exponential distribution of the free
volume, which is well confirmed by experiments with dry granulates
\cite{Aste2005}. The conditions (\ref{DensityCondition}) and
(\ref{JammingOfA}) determine the A-neighbor distribution uniquely:
\begin{eqnarray}
P_{\text{A}}(s) \ \D\,\text{vol}(s) =
\frac{D/(\sigma_D d^D)}{{\phi_{\text{J}}} / {\phi} - 1} \ \exp \left(
-\frac{\left(1+\frac{s}{d} \right)^D -1 }{{\phi_{\text{J}}} /
{\phi} - 1} \right) \ \D\,\text{vol}(s) \ .
\label{FreeVolumeTheory1}
\end{eqnarray}
The volume element for $D=2$ is
\begin{eqnarray}
\D\,\text{vol}(s)=\sigma_D \; r^{{(D-1)}} \; \D r= \pi d \;
(1+s/d) \; \D s\ .
\end{eqnarray}
The contribution $g_{\text{A}}$ which A-neighbors give to the pair
correlation is equal to the A-neighbor
distribution $P_{\text{A}}$ (\ref{FreeVolumeTheory1}) up to a prefactor, so that
\begin{eqnarray}
g_{\text{A}}(s)=\gc \ \exp \left( -\frac{\left(1+\frac{s}{d} \right)^D -1
}{{\phi_{\text{J}}} / {\phi} - 1} \right) \label{FreeVolumeTheory2}
\end{eqnarray}
is determined as soon as we know the athermal contact value, $\gc=g_{\text{A}}(0)$.
This contact value follows from the classical free volume theory \cite{SalsburgWood1962}
(which was based on \cite{Hirschfelder1950}),
\begin{eqnarray}
\frac{P}{n T} = \frac{D}{{\phi_{\text{J}}}/{\phi}-1} + {\cal O}(1) \ , \label{FreeVolumeTheory3}
\end{eqnarray}
in conjunction with the general relation between the particle-wall correlation
$\gcwall$ and the pair correlation $\gc$,
\begin{eqnarray}
\frac{P}{n T} = \gcwall = 1 + 2^{D-1} \phi \ \gc\ . \label{gpwgpp}
\end{eqnarray}
As a consequence, we obtain
\begin{eqnarray}
\frac{2^{D-1}}{D} \phi \ \gc = \frac{1}{{\phi_{\text{J}}}/{\phi}-1}  \label{FreeVolumeGC}
\end{eqnarray}
close to jamming.
Expression (\ref{FreeVolumeGC}) is exact for $D=1$, and has been
confirmed as the asymptotic behavior of the diverging pressure
close to jamming for $D=2$ \cite{LudingZustandsgleichung,
LudingTransport} in event-driven simulation with accuracy
$10^{-4}$. We remark that this expression is not limited to weak
polydispersity and
has been confirmed for polydispersity
far above $0.1$ in the glass state
\cite{Speedy1994, Donev2006}.


Inserting (\ref{FreeVolumeGC}) in (\ref{FreeVolumeTheory2}), we
have as our first central result a closed expression for the
near-contact pair correlation of neighbors which form exact
contacts in the jamming limit (so-called A-neighbors):
\begin{eqnarray}
g_{\text{A}}(s) = \gc \exp{\left( - \frac{2^{D-1}}{D} \ \phi \ \gc \
\left[\left(1+\frac{s}{d} \right)^D -1 \right] \right)} \ .
\label{ShortRangeCorr}
\end{eqnarray}
Eq.~(\ref{ShortRangeCorr}) implies for the derivative at contact,
\begin{eqnarray}
d \ g'_{\text{A}}(0) = - 2^{D-1} \phi \ g^2_{\text{A}}(0) \ ,
\label{FormulaDenseDerivative}
\end{eqnarray}
a quadratic dependence on the contact value
$\gc=g_{\text{A}}(0)$. Eq.~(\ref{FormulaDenseDerivative})
can be viewed as a consequence of normalization: the height of the
contact peak is $\gc$ and so the width is of the order
$1/\gc$, which means that the negative slope is of the
order $g^2_{\text{c}}$. In fact, writing the A-neighbor
correlation function $g_{\text{A}}(s)$ in terms of the contact value
$\gc$, as we did in (\ref{ShortRangeCorr}), is the
natural form to express the density dependence of $g_{\text{A}}$
because this manifests that the coordination number of A-neighbors
is density independent:
\begin{eqnarray}
K_{\text{A}} &=& n  \int g_{\text{A}}  \ \D\, \text{vol}  \nonumber \\
&=&\frac{2^D \ D}{d} \ \phi \ \gc \ \int_{s=0}^\infty \exp{\left( - \frac{2^{D-1}}{D} \ \phi \ \gc \
\left[\left(1+\frac{s}{d} \right)^D -1 \right] \right)} \ \left(1+\frac{s}{d}\right)^{D-1} \ \D s \nonumber  \\
&=& 2D   \ . \label{IsostaticResult}
\end{eqnarray}
More significantly, $K_{\text{A}}$ equals exactly the isostatic
contact value $2D$, which is obviously correct for particles on a
line ($D=1$) and is the accepted value for ideal discs and spheres
in $D=2$ and $D=3$ dimensions respectively \cite{Donev2005,
Donev2007, Sperl2007}.
The finding (\ref{IsostaticResult}) is an essential confirmation
of consistency of our approach, since it is independent from
conventional arguments based on the rank of the rigidity matrix
(which accounts for global constraints on the degrees of freedom)
\cite{Donev2005}.

As the contact value $\gc$ (\ref{FreeVolumeGC}) grows to
infinity in the jamming limit, $\phi \rightarrow \phi_{\text{J}}$,
the constant integral (\ref{IsostaticResult}) implies that $n \
g_{\text{A}}(s)$ becomes a delta distribution with 'weight' $2D$ at contact, $s=0$.


\subsubsection{The Background Contribution: The B-Neighbors}
The configuration space is spanned by all particle positions $\{{\bf r}_i\}$.
Consequently, a jammed configuration is -- aside from a small fraction of rattlers
\footnote{Typically one or two percent for packings produced by
the Lubachevsky-Stillinger algorithm (Lubachevsky \cite{Lubachevsky} for 2D,
\cite{Donev2005} for 3D)}
-- an isolated configuration point, and the set of jammed
configuration is a set of discrete points. When the density is
slightly relaxed, a finite system remains confined to a finite
environment around the jamming point (cf. \cite{ConnellyLecture},
p.~35). As density is lowered further, these environments are no
longer isolated so that the system is able to migrate between
theses 'islands of jamming'.


The stability analysis of contact networks
\cite{Roux2000,Connelly2005Math,Connelly2005IHP,Donev2005,Donev2004}
has put forth the result
that frictionless spheres (except for the singular limiting case
of a monodisperse crystal) jam strictly in an isostatic packing with $2D$
contacts per particle on average, as confirmed numerically
\cite{Donev2005} for $D=3$, be the state random (glass regime in Fig.~\ref{chpEqStGraphic2b})
or locally ordered. Therefore we can identify within an island of jamming on average
four neighboring particles in $D=2$ dimensions which are close to
the reference particle,
and which will be in contact with the reference particle, $\sA
\rightarrow 0$, in the jamming limit, $\phi \rightarrow
\phi_{\text{J}}$. These are the A-neighbors with the contribution $g_{\text{A}}$ to the pair correlation
derived in (\ref{ShortRangeCorr}).
Furthermore, it is a mathematical fact that
any discrete set of points in flat two-dimensional space has on
average six Delaunay/Vorono\"i neighbors \cite{Meijering1953},
two of which have no contact to the reference particle, $g_{\text{B}}(0)=0$.
Hence, on the mean
field level the following picture arises: Beside the four
A-neighbors there are two B-neighbors which are sterically
hindered by other particles from further approach to the reference
particle. Summing up the contributions of A- and B-neighbors,
\begin{eqnarray}
g^{\text{dense}}(s)=g_{\text{A}}(s)+g_{\text{B}}(s) \label{DecompPairCorr} \ ,
\end{eqnarray}
gives us the pair correlation function near contact.
\begin{figure} \begin{center}
\epsffile{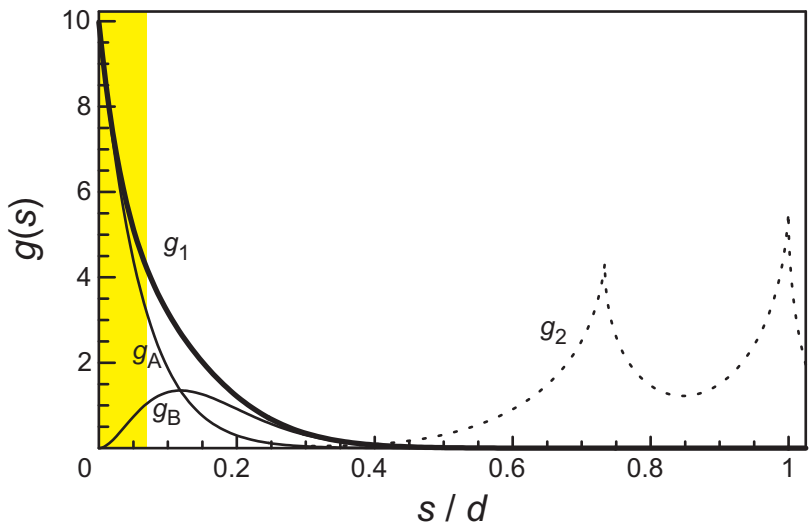} \caption{The pair correlation near contact
resulting from free-volume considerations.
Close to jamming we distinguish between neighbors which form exact
contacts in the jamming limit (contribution $\gA$, Eq.~(\ref{ShortRangeCorr}))
and those that are blocked at positive separation $s$
(curve $\gB$, Eq.~(\ref{ShortRangeCorrB})).
The near-contact correlation is the sum of both contributions.
For this plot the density is chosen to $\phi=0.8$. The dashed curve sketches
a typical second shell consisting of the second Vorono\"i neighbors.
They are out of the interaction range, $0<s<\scrit$,
which is indicated by the highlighted stripe.}
\label{chpEqStgAgB}
\end{center} \end{figure}
The pair correlation near contact which arises from these blocked
states, $g_{\text{B}}$, is discussed in detail in appendix~
\ref{AppendixgB2D}. The essential result is that the configuration
space of blocked states tends quadratically to zero in $\sB$, so
that to leading order the normalization of two B-neighbors for
$D=2$ determines the B-contribution in (\ref{DecompPairCorr}):
\begin{eqnarray}
g_{\text{B}} (\sB) &=& {\cal N} \ P_{\text{B}}(\sB) \nonumber \\
 &=& \frac{1}{\phi c^3_{\text{B}} } \ \left(\frac{\sB}{d}\right)^2 \
 \Exp{-\left[\left(1+\frac{s}{d} \right)^2 -1 \right]/c_{\text{B}}}  \left[1 + {\cal
 O}\left(\frac{\sB}{d}\right)\right] \label{ShortRangeCorrB}
\end{eqnarray}
with $c_{\text{B}} = \phi_{\text{max}}/\phi-1$. In Fig.~\ref{chpEqStgAgB} the resulting
near-contact pair correlation \ref{DecompPairCorr} for the dense regime is
shown as the sum of $\gA$ and $\gB$.


\subsection{The Dilute and Moderately Dense Regime} \label{SecDeplForce}
\begin{figure} \begin{center}
\epsffile{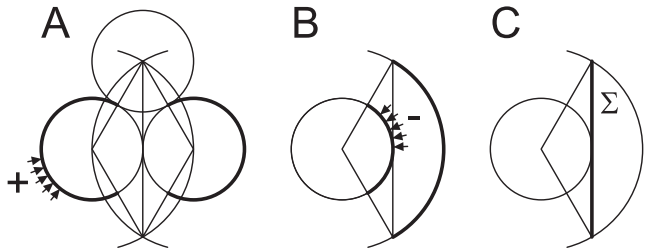} \caption{Origin of the depletion
force attracting neighboring particles that are separated by less
than a particle diameter. One may either think of this as an
entropic force, due to the decrease of excluded volume when the
shells of excluded volume overlap. Equivalently one may view this
as the net force due to isotropic bombardment. Obviously, the
integration over the solid arc in A is up to a sign equivalent to
the integration in plot B. In B the integration is over the outer
solid arc, which is the configuration space of the third
particle's coordinate at impact. Since the integration in B is
projected by a $\cos$-factor to give the axial symmetric force
component, we can equivalently drop the $\cos$-factor and
integrate over the submanifold indicated by the
solid line $\Sigma$ in C.}
\label{chpEqStGraphic5}
\end{center} \end{figure}

In this part we turn to the free rheological regime, $0 < \phi <
\phi_{\text{o}}$. When two spheres are closer than one diameter,
$s<d$, they shield each other from certain collisions events. If
one was to neglect three-particle correlations, the isotropic
bombardment by `third' particles gives rise to the well-known
attractive depletion force first proposed by S.~Asakura and
F.~Oosawa \cite{Asakura1, Asakura2}. As is evident from
Fig.~\ref{chpEqStGraphic5}, summing up equal contributions over
the accessible cross section is equivalent to the pressure exerted
onto the submanifold indicated by the solid line in
Fig.~\ref{chpEqStGraphic5}C and denoted by $\Sigma$.

This depletion force, as well as the liquid bridge force which we
will take into account in the next section, will affect the pair
correlation function. A systematic way to study this effect has
been worked out by Hansen \emph{et al.} \cite{Piasecki}, resulting
in a Fokker-Planck equation for the two-particle distribution
function. After integrating out the momenta and the center of mass
coordinates, one finds that the depletion force as well as
other non-entropic pair forces (such as the liquid bridge force),
give rise to a Boltzmann factor,
\begin{eqnarray}
 g(s) \propto \exp \left(-\frac{V(s)}{T}\right) \ . \label{BoltzmannFactor}
\end{eqnarray}
For the depletion force
\begin{eqnarray}
F_{\text{depl}}=-V'_{\text{depl}}=T \ n \ \gc \ \Sigma \
, \label{GenDeplForce}
\end{eqnarray}
where
\begin{eqnarray}
n \ \gc \ \Sigma \ \D s = \frac{\D V_{\text{conf}}}{V_{\text{conf}}} = - \D \ln g(s) \label{Depl=LogDiv}
\end{eqnarray}
is the infinitesimal logarithmic change of the excluded area
(or the configuration space per particle,
$V_{\text{conf}}$), when the particles are separated by $s<d$, and $\Sigma$ denotes
the size of the corresponding section (line or area) in Fig.~\ref{chpEqStGraphic5}C.
At contact, $s=0$, the size of the integration section
$\Sigma$ is
\begin{eqnarray}
\Sigma = \frac{\sigma_{D-1}}{D-1}
\left(\frac{\sqrt{3}}{2}d\right)^{D-1} \ , \label{OACrossSection}
\end{eqnarray}
which yields $V_{\text{depl}}$ to leading order in $s$. The depletion
effect with the potential
\begin{eqnarray}
\frac{V_{\text{depl}}}{T} = \frac{9}{2} \ \phi \ \gc \
\frac{s}{d} \left(1 -
\frac{s}{3d}-\left(\frac{s}{3d}\right)^2\right)
\end{eqnarray}
for $D=3$ has been confirmed in \cite{Mason1992} by computer simulations.
Polydispersity is known to have a minor effect on the depletion
attraction \cite{GouldingHansen2001}. For $D=1$
(\ref{Depl=LogDiv}) gives the Poisson distribution
${V_{\text{depl}}}/{T} = \phi \ \gc \ {s}/{d}$
which is exact only for $D=1$.

In two dimensions, the depletion potential is
\begin{eqnarray}
\frac{V_{\text{depl}}}{T} & = &  \frac{2}{\pi} \ \phi \
\gc \ \left( 4 \arctan \frac{\gvol}{W} +
\gvol W
- C \right) \label{DeplExponentVoll} \\
& = & 2\gc \frac{\phi}{\phi_{\text{max}}} \frac{s}{d} +
{\cal O}\left(\frac{s}{d}\right)^2
\label{DeplExponent}
\end{eqnarray}
for $D=2$ with $\gvol(s)={1+s/d}$, the square root
$W(s)=\sqrt{(1-s/d) (3+s/d)}$ and the constant $C=2 \pi/3 +
\sqrt{3}$ to have $V_{\text{depl}}=0$ at $s=0$. The first line
(\ref{DeplExponentVoll}) is valid for $0\leq s\leq d$, and the
second line (\ref{DeplExponent}) suffices for the region of
interest, $0\leq s\leq \scrit\ll d$. For the application
of results on the near-contact decay of the pair correlation
function, such as (\ref{DeplExponent}),
we prefer the exponential notation (used before in the dense case
(\ref{ShortRangeCorr})) because it is most elegant to perform
volume integration:
\begin{eqnarray}
g^{\text{dilute}}_{\text{AO}}(s) = \gc \exp{\left(
-\frac{\phi}{\phi_{\text{max}}} \ \gc \
\left[\left(1+\frac{s}{d} \right)^2 -1 \right] \right)} \left[1 +
{\cal O}\left(\frac{\sB}{d}\right)^2\right]
\label{ShortRangeCorrDeplAO}
\end{eqnarray}
for $D=2$.
In this notation the dilute and dense behavior of the pair
correlation are conveniently compared, showing that the result
(\ref{ShortRangeCorrDeplAO}) for the gaseous/fluid regime differs
by the factor $1/\phi_{\text{max}}=1.10$ in the exponent from the
dense result (\ref{ShortRangeCorr}) close to jamming, so that
according to (\ref{ShortRangeCorrDeplAO}) the depletion force
falls-off slower than the configuration density $\phi
\gc$. We will now show that this is due to an
over-estimation of the depletion force, caused by neglecting
correlated three-particle events: when the plane of incidence of
the third particle closely coincides with the symmetry plane
$\Sigma$, the incoming particle will hit in short sequence the
pair of particles considered, which increases very effectively the
exchange of momentum, i.e. the depletion attraction is reduced.

\begin{figure} \begin{center}
\epsffile{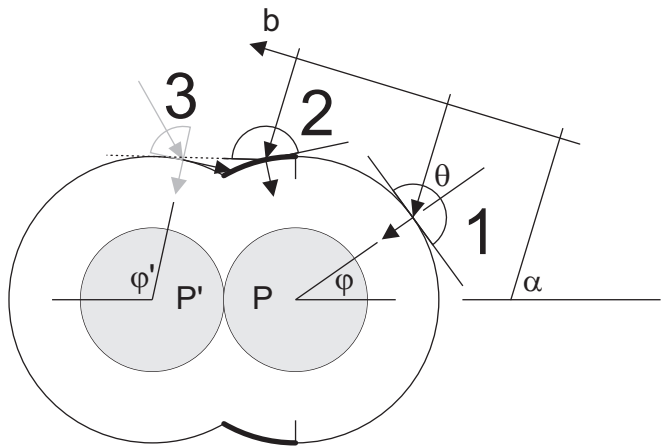} \caption{Three
contributions to the effective force between a pair of particles P
and P'. The collisions events 1 are attractive, while the events 2
cause a weaker repulsive forces. Furthermore the attraction is
weakened by the temporally correlated collisions events 3.}
\label{chpEqStThreeParticleEvents}
\end{center} \end{figure}

\begin{figure} \begin{center}
\epsffile{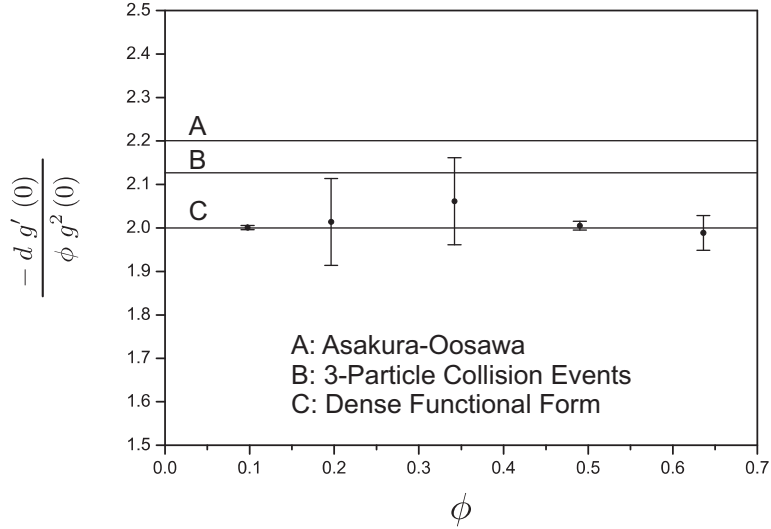} \caption{Functional test of the
near-contact pair correlation (\ref{ShortRangeCorrDepl}). The
vertical axis is proportional to the depletion force,
$F_{\text{depl}}\propto -{\D \ln g} / {\D s}$ at contact, divided
by the configuration density $\phi \ \gc$. This fraction $Z$ is
predicted to be density independent by (\ref{ShortRangeCorrDepl})
and to assume the value $Z=2$ (line C). Line A corresponds to the classical Asakura-Oosawa
result, which is only valid for large beads immersed in a bath of small beads.
In line B the corrections due to temporally correlated
collisions events (derived in the text) have been taken into account.
These events occur when a third particle of equal size
strikes a pair of particles with a given separation $s \ll d$ as sketched in
Fig.~\ref{chpEqStThreeParticleEvents}. We proceed using the value $Z=2$ (line C)
because it agrees best with the simulation. Furthermore, $Z=2$ corresponds to a
near contact correlation function which is of exactly
the same form as the function $\gA$ we use in the dense case, when expressed
in terms of the configuration density $\phi \ \gc(\phi)$.} \label{chpEqStTestDepl}
\end{center} \end{figure}

To determine analytically and numerically the effect of correlated
collisions which correct the Asakura-Oosawa result
(\ref{ShortRangeCorrDeplAO}) we define the dimensionless measure
\begin{eqnarray}
Z=\frac{4}{\pi} \frac{F_{\text{depl}}}{n T d
\gc}=-\frac{d g'_{\text{c}}}{\phi \ g^2_{\text{c}}} \ , \label{DefinitionZ}
\end{eqnarray}
for which the Asakura-Oosawa approach (\ref{GenDeplForce}) and
(\ref{OACrossSection}) gives
$Z_{\text{AO}}=\frac{4}{\pi}\sqrt{3}\approx2.205$ (Line A in
Fig.~\ref{chpEqStTestDepl}). When we take correlated
three-particle events into account, there are three contribution.
Firstly, an attractive contribution $Z_1>0$ due to collisions on
the front side of the pair, indicated by '1' in
Fig.~\ref{chpEqStThreeParticleEvents}, which fall in the range
$-\pi/2<\varphi < \pi/2$. The corresponding value $Z_1$ is easily
integrated. Isotropy of the state demands that the angle $\alpha$
between the symmetry axis of the pair $P'P$ and the incoming
momentum ${\bf p}_{\text{i}}$ is uniformly distributed, as well as
the impact parameter $b$ (cf.
Fig.~\ref{chpEqStThreeParticleEvents}). These collision parameters
are related by $(\alpha,b)=(\varphi+\theta,d \sin \theta)$ to the
position $\varphi$ on P and the angle of incidence $\theta$ with
respect to the normal of P, which implies that $\varphi$ is
uniformly distributed and $\theta$ is weighted by the
cosine-factor $\cos \theta$. Integration over $-\pi/2<\varphi <
\pi/2$ yields the axial force contribution
\begin{eqnarray}
F_{1}= 2 T n \gc d \ ,
\end{eqnarray}
so that $Z_1=8/\pi\approx 2.546$.

Secondly, the attraction is weakened by collisions hitting P in
the remaining range $\pi/2<\vert \varphi \vert <
\varphi_{\text{max}}(s)$ (which we refer to as the 'broad side')
giving rise to $Z_2<0$. At contact $\varphi_{\text{max}}(s=0)$ is
$2\pi/3$. For these collisions the incidence is shadowed by the
partner particle P' so that the angle of incidence $\theta$ is
restricted to $-\pi/2 < \theta < \theta_{\text{max}}(\varphi)$.
(Confer the collision event '2' in
Fig.~\ref{chpEqStThreeParticleEvents}.) Some trigonometry
determines $\theta_{\text{max}}(s,\varphi)$ by the relation
\begin{eqnarray}
\left(1+\gvol(s) \cos \varphi\right) \sin
\theta_{\text{max}} = 1-
\gvol(s)\cos\theta_{\text{max}} \sin \varphi \ ,
\end{eqnarray}
which allows for an explicit function of $\varphi$ at $s=0$:
\begin{eqnarray}
2 \cos \theta_{\text{max}}(\varphi) = \tan
\frac{\varphi}{2}-\sqrt{1+2 \cos \varphi} \ .
\end{eqnarray}
After integrating over the impact momenta $p_{\text{i}}$ in the
rest frame of P, the axial force imposed on P is
\begin{eqnarray}
F_{2}(s)=\frac{4}{\pi} \ T n \gc d \
\int_{\pi/2}^{\varphi_{\text{max}}(s)} \ \D \varphi \ \cos \varphi
\ \int_{-\pi/2}^{\theta_{\text{max}}(\varphi)} \ \D \theta
\cos^2\theta \ ,\label{chpEqStF2}
\end{eqnarray}
where the $\cos \varphi$ projects the force on the symmetry axis
of the pair PP'. The $\cos\theta$ factor appears quadratically in
the integrand (\ref{chpEqStF2}) because of the cosine-distribution
(or equivalent, because the Enskog collision frequency is
proportional to the radial velocity $(p_{\text{i}}/m) \cos \theta$),
and the transferred momentum which is $p_{\text{i}} \cos \theta$.
Symmetry allows us to integrate over the upper half, $\pi/2< \varphi <
\varphi_{\text{max}}(s)$ in (\ref{chpEqStF2}) and multiply by 2
with the general result
\begin{eqnarray}
Z_{2}(s)=\frac{8}{\pi} \int_{\pi/2}^{\varphi_{\text{max}(s)}} \ \D
\varphi \ \cos \varphi \left[ \frac{1}{2} +
\frac{\theta_{\text{max}}(s,\varphi)}{\pi} + \frac{\sin 2
\theta_{\text{max}}(s,\varphi)}{2 \pi} \right] \label{chpEqStZ2} \
,
\end{eqnarray}
and the numerical value $Z_2(0) = -0.32813(9)$.

Thirdly, the most obvious and important correction on the
three-particle level comes from double collisions denoted by 3 in
Fig.~\ref{chpEqStThreeParticleEvents}. The third particle hits
first P' (gray arrow in Fig.~\ref{chpEqStThreeParticleEvents})
from the broad side at $\varphi'\in(\pi/2,\theta_{\text{max}})$.
The radial component $p_{\text{i}} \cos\theta'$ of its incoming
momentum $p_{\text{i}}$ is transferred to P', which is why the
third particle moves on tangentially to the circular cross section
of P' with momentum $p_{\text{i}} \sin\theta'$ to collide shortly
afterwards with particle P. Here the momentum transferred is the
radial component with respect to P, $p_{\text{i}} \sin\theta' \cos
\theta$, so that
\begin{eqnarray}
F_3(s)=\frac{4}{\pi} \ T n \gc d
\int_{\pi/2}^{\varphi_{\text{max}}(s)} \ \D\varphi' \ \cos
\varphi(\varphi') \ \cos \theta(s,\varphi') \
\int_{0}^{\theta_{\text{max}}(s,\varphi')} \D \theta' \
\cos\theta' \sin\theta' \ .
\end{eqnarray}
The collision point on P' described by $\varphi'(\varphi)$ is
related to $\varphi$ (the subsequent collision point on P) by
$\cos(\varphi'-\varphi)=1+\gvol(s) \cos \varphi$.
The incident angle $\theta$ on P is independent of $\theta'$ and
given by $\sin \theta(s,\varphi')=1+\gvol(s) \cos
\varphi'$. After the elementary $\theta'$-integration we find
\begin{eqnarray}
Z_3(s)=\frac{8}{\pi^2} \ \int_{\pi/2}^{\varphi_{\text{max}}(s)} \
\D\varphi' \ \cos \varphi(\varphi') \ \cos \theta(s,\varphi') \
 \sin^2\theta_{\text{max}}(s,\varphi') \ ,
\end{eqnarray}
and $Z_3(0)=-0.091593(7)$. Summing up the three contributions
gives $Z_{\text{corr}}=\sum_{i=1}^3 Z_i \approx 2.127$ which is
shown as the line B in Fig.~\ref{chpEqStTestDepl}.

Based on our numerical data shown in Fig.~\ref{chpEqStTestDepl},
we shall in the sequel assume the value
\begin{eqnarray}
Z_{\text{sim}}=2 \ . \label{Z=2}
\end{eqnarray}
By virtue of good statistics the simulation at
$\phi=0.097$ gave $Z_{\text{sim}}=2.0009 \pm 0.0050$, and
Fig.~\ref{chpEqStTestDepl} suggest this result to hold with few percent
limits very well over the entire density regime
$0<\phi<\phi_{\text{o}}$ considered in this
subsection. The value $Z=2$ determines the near-contact pair
correlation uniquely to be
\begin{eqnarray}
g^{\text{dilute}}(s) = \gc \exp{\left( -{\phi} \
\gc \ \left[\left(1+\frac{s}{d} \right)^2 -1 \right]
\right)} \left[1 + {\cal O}\left(\frac{\sB}{d}\right)^2\right]
\label{ShortRangeCorrDepl}
\end{eqnarray}
for $D=2$. Satisfactorily, this result (\ref{ShortRangeCorrDepl}) has
exactly the same functional dependence on the configuration
density $\phi \ \gc$ as the formula put forward for the
dense case (\ref{ShortRangeCorr}) in the previous subsection.
While three-particle collisions are obviously important since they
shift $Z$ in the right direction, no analytic explanation for this
coincidence corresponding to the value $Z=2$ is provided at
present. Yet we shall see in the next section that any value other
than $Z=2$ would lead to inconsistencies when we introduce the
liquid bridge interaction.

We finally remark that the result (\ref{ShortRangeCorrDepl})
strongly differs from the 'Poissonian fluid' \cite{EdgalHuber1993},
for which
the contact correlation $\gc-1>0$ is ignored.
Even at the lowest density ($\phi=0.1$) considered in
Fig.~\ref{chpEqStTestDepl} the Poisson fluid
would give $Z_{\text{Poisson}}(\phi=0.1)=1.7$ which is 15\%
below the simulation value,
and the deviation from $Z=2$ grows with density $\phi$.

\section{The Pair Correlation under the Hysteretic Interaction}
In this section we
dress up the pair correlation function in order to describe
the status of the liquid-bonds which are created and ruptured hysteretically
in wet granular matter.
We will proceed in two steps: first, we introduce in part
\ref{ForcelessBridges} the liquid
bridges as hysteretic but forceless objects which follow
the unperturbed particle dynamics. As a result, a
direct relation of the dynamical system and the limiting case of
isostatic granular packings \cite{Donev2004, Sperl2007} at rest is found.
In \ref{BridgesWithForce} we turn on the liquid bridge force to
its physical value, so that the bridges
unfold their back-reaction on the granular dynamics. In the limit of
low granular temperatures, $T\ll \Ecb$, the particles stick together.
For this frozen state of wet granular matter the bridge coordination $K$
is computed analytically as a function of the rupture length $\scrit$, and
we find very good agreement with simulations.

\subsection{The Hysteretic Coupling} \label{ForcelessBridges}
\begin{figure} \begin{center}
\epsffile{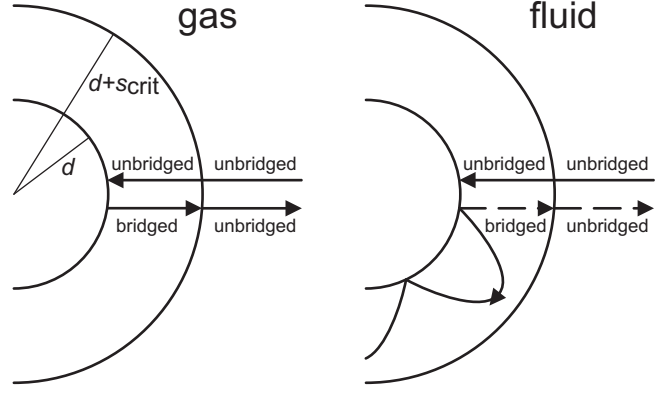} \caption{The hysteretic
interaction in a wet granular gas or fluid can either lead to
scattering or bound states. Note that the formation and rupture of
the liquid bridge is spatially separated, which gives rise to
a hysteretic loss and a coupling between the pair correlation
functions $g^{\text{b}}$ for neighbors with and without, $g^{\text{u}}$,
capillary bridge.
In this sketch the maximal liquid bridge length
$\scrit$ is drawn largely exaggerated. For a typical
volume fraction of 1\% wetting liquid added to the volume of jammed granular matter
one finds $\scrit/d \approx 0.07$ \cite{Herm05}.} \label{chpEqStGraphic6}
\end{center} \end{figure}

Due to the hysteretic interaction, the pair correlation $g$
is no longer a function of the particle separation $s$. In order to
include the knowledge about the collision history the
configuration space has to be enlarged in two respects:
Obviously we distinguish between pairs with and without
liquid bridges, which we denote by superscript indices,
$g^{\text{b}}(s)$ and $g^{\text{u}}(s)$ respectively, for `bridged' and `unbridged'
neighbors (cf. Fig.~\ref{chpEqStGraphic6}).
In addition, time reversal-symmetry is broken
by the formation of the capillary bridge at contact. Hence
we distinguish approaching pairs (with a negative
relative velocity) which might collide and form a liquid bridge in
the future, and those that move apart so that they can rupture the
liquid-bond in the future. This relative velocity is denoted by a subscript arrow.

As we discuss the radial pair distribution, contact and rupture
become the important points on the $s$-axis of the pair
correlation function. At theses points the functions
$g^{\text{b}}$ and $g^{\text{u}}$ are coupled according to the
hysteretic transition of the bond status.
We use an intuitive notation to refer to these points:
\begin{eqnarray*}
\gunr &=&  \left\{\begin{minipage}{0.3\textwidth}
The probability for a pair\\
at rupture distance \\
approaching without bridge.
\end{minipage} \right. \\
\gbpc &=& \left\{\begin{minipage}{0.3\textwidth}
The probability for a pair \\
at contact \\
moving away with bridge.
\end{minipage} \right. \\
\text{etc.} &&
\end{eqnarray*}
The contact point $s=0$ (or to be more precise: the right-sided
limit $s=0+$) is denoted by a circled $\Hc$, and the rupture at
$s=\scrit$ by the circled $\Hr$. Infinitesimally close to contact, there are
four detailed correlation values: the bridge-connected and the unconnected states,
either particularized by the sign of the relative velocity. The same is
true for the left-sided limit $s=\scrit-$ of the rupture
point. An infinitesimal distance beyond this point, at $s=\scrit+$,
there is only the unbound state possible with the two signs for
incoming and outgoing velocities. This gives us in total ten
detailed pair correlation coefficients. These are determined by
the following ten equations describing the hysteretic flow of
probability, as it we can be read off from
Fig.~\ref{chpEqStGraphic6}:
\begin{eqnarray}
&&\text{\emph{Conditions on the contact shell}} \nonumber  \\
\gupc &=& 0                                     \label{E1} \\
\gbpc &=& \gunc + \gbnc                         \label{E2}
\end{eqnarray}
The Eq.~(\ref{E1}) expresses that no particles rebound without a
liquid-bond, but rather that all return with a bridge as stated by
(\ref{E2}). This implies that the particle number is conserved in
collisions (in contrast to the absorbent dynamics modelled in
\cite{Vasily2006} for $D=1$).
\begin{eqnarray}
&&\text{\emph{Domain of capillary interaction}} \nonumber \\
\gbr  &=& \bs(\scrit) \ \gbc                                            \label{E4} \\
\gur  &=& \us(\scrit) \ \guc                                            \label{E5} \\
\gupr &=& \gamma_{\text{pass}} \ \gunr   \label{E6}
\end{eqnarray}
The functions $\us(s)$ and $\bs(s)$ take into account the near-contact
decay of the pair correlation without and with liquid bond,
respectively.
The last Eq.~(\ref{E6}) describes spectator grains, i.e. grains
which pass through the domain of possible
capillary interaction without bridge formation. The fraction of these passing particles,
$\gamma_{\text{pass}} = 1-(1+\scrit/d)^{1-D}$ ($=1/{\left(1+d/\scrit\right)}$
for $D=2$) equals the gap between the considered cross section
$(2d+2\scrit)^{D-1}$ of the capillary
interaction and the hard-core cross section $(2d)^{D}$.
\begin{eqnarray}
&&\text{\emph{Conditions on the rupture shell}} \nonumber \\
\gbnr &=& 0                             \label{E7}        \\
\gunr &=& \gonr                         \label{E8}        \\
\gbpr + \gupr &=& \gopr                 \label{E9}
\end{eqnarray}
The Eqs.~(\ref{E7}, \ref{E8}) state that only unbound particles
enter the domain of capillary interaction, and (\ref{E9})
describes the rupture of a capillary bridge when the pair escapes
from the domain.

The hysteretic capillary dynamics is coupled to
the hard particle dynamics by the source term of new unbound pairs of
particles entering the capillary interaction range:
\begin{eqnarray}
&&\text{\emph{Source Term}} \nonumber \\
\gunc + \gunr / \us(\scrit) &=& ({1-K/K_{\text{sites}}})
\gcHSS \label{E3}
\end{eqnarray}
The left-hand side is the current of approaching unbound neighbors
(measured at contact). If all neighbors were unconnected, $K=0$,
this current would equal the dry value $\gcHSS$. But since there
are $K$ neighbors with bonds out of the $K_{\text{sites}}$
'docking sites' which are sterically accessible for liquid bonds,
the remaining unconnected fraction is $1-K/K_{\text{sites}}$.

The final tenth equation is the stationary state condition, which
demands that the rupture frequency equals the binding frequency:
\begin{eqnarray}
&&\text{\emph{Stationary state condition}} \nonumber \\
f_{\text{bind}} &=& f_{\text{rupt}}  \ . \label{E10}
\end{eqnarray}

These frequencies follow from the probability to have a particle
on the collision or rupture shell, respectively, multiplied by the
radial component of the relative velocity under the condition that
the particle moves in the appropriate direction for the event to
occur. This is analogous to the case $D=1$ \cite{Fingerle2006},
with the only difference that here we have to integrate over
shells:
\begin{eqnarray}
f_{\text{bind}} &=& 2^{D+1} D \ \sqrt{\frac{T}{\pi}} \ \frac{\phi}{d} \ \gunc \text{, and} \label{fbind} \\
f_{\text{rupt}} &=& 2^{D+1} D \ \sqrt{\frac{T}{\pi}} \ \frac{\phi}{d} \ \gbpr \ \gvol(\scrit)   \ .         \label{frupt}
\end{eqnarray}
The volume factor $\gvol(s)=(1+s/d)^{(D-1)}$ takes
the increased size of the outer rupture shell as compared to the
inner binding shell into account.

Eliminating those correlation coefficients that are identically
zero (\ref{E1}, \ref{E7}), we can arrange the coupling equations
for the domain of capillary interaction as a $6 \times 6$ matrix
system:
\begin{equation}
\begin{array}{l}
\text{Collision:}\\
\text{With Bridge:}\\
\text{Unconnected:}\\
\text{Stationarity:}\\
\text{Spectators:}\\
\text{Source:}
\end{array}
\left(\begin{array}{cccccc}
1&1&-1&0&0&0\\
0&\bs&\bs&0&-1&0\\
\us&0&0&-1&0&-1\\
-1&0&0&0&\gvol&0\\
0&0&0&-1&0&\gamma_{\text{pass}}\\
1&0&0&0&0&1/\us
\end{array} \right)
\circ
\left(\begin{array}{c}
\gunc \\
\gbnc \\
\gbpc \\
\gupr \\
\gbpr \\
\gunr
\end{array}\right)
= ({1-K/K_{\text{sites}}}) \ \gcHSS \left(
\begin{array}{c}
0\\
0\\
0\\
0\\
0\\
1
\end{array}\right) \label{HystereticSystem}
\end{equation}
The $\gamma$-functions in the matrix are to be evaluated at
$s=\scrit$. As it has to be on physical grounds, this
system is non-singular with determinant
$(1+\scrit/d)^{D-1} \ (2+\gamma_{\text{pass}}) \ \bs(\scrit)
>0$.

\begin{figure} \begin{center}
\scalebox{0.5}{\epsffile{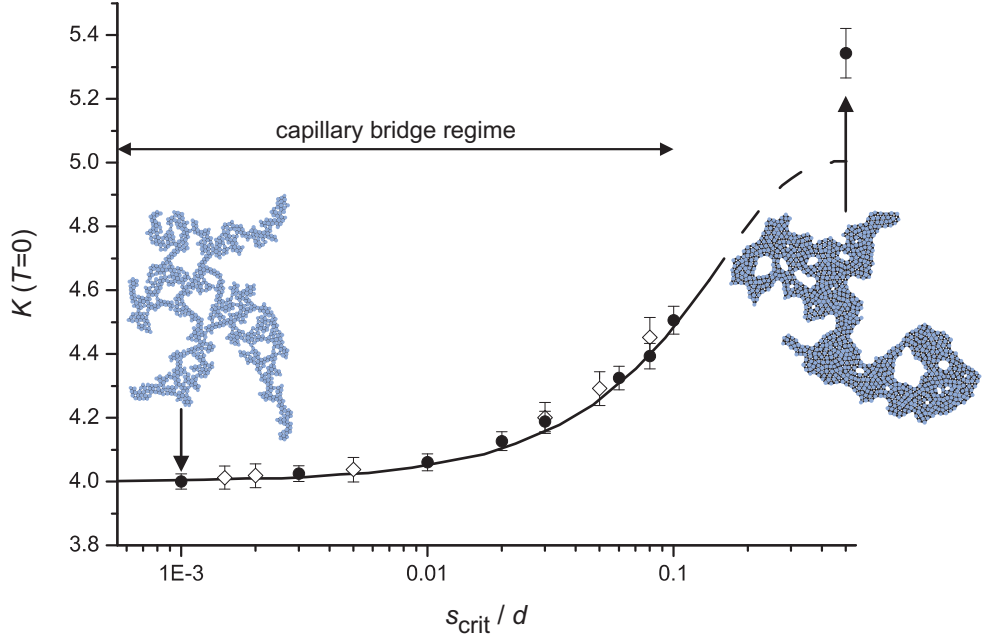}}
\caption{The capillary bridge coordination $K$ in the low temperature limit,
$T\ll \Ecb$. As proven in the text, $K$
converges to the athermal function
$K_{\text{sites}}(\phi,\scrit)$ in this low temperature
limit. The solid line is
$K_{\text{sites}}(\phi_{\text{o}},\scrit)$ over a wide
range of maximal bridge lengths $\scrit$. Points
represent final states of free cooling simulations with 1000
particles of uniformly distributed polydispersity
$\Delta d = 0.06 d$. The open symbols are clusters
with winding number one (cylindrical topology), connected over one periodic boundary on a
rectangular domain. Such structures have internal tensile strength
which necessitates a slightly increased coordination, visible as a
small shift compared to the closed symbols which represent
localized clusters (as the two examples drawn in the plot). As predicted by
Eq.~(\ref{ExactContactLimitD=2}) of the presented theory, the structures emerging
with exact contacts, $\scrit \rightarrow 0$, are found to be precisely isostatic,
$K_{\text{sites}}=4$.
The line is the the analytic result (\ref{ExactContactLimit}), for
which very good agreement is found with the simulations over the
entire range of the capillary bridge regime, $0 < s < 0.2 r$ (with
$r$ the particle radius), which is indicated in the figure. Beyond
this regime, the theory does not hold because in the derivation we
limited ourselves to the leading order in $\scrit/d$. More
importantly, the rupture length $\scrit$ cannot be further
increased beyond the capillary regime by simply increasing the
liquid content in the granular sample. As mentioned in the
introduction, liquid bridges residing on the same sphere would
rather merge \cite{MarioNatureMaterials} into more complicated
objects.} \label{chpEqStConfirmKsites}
\end{center} \end{figure}

The last row of the system (\ref{HystereticSystem}) describes the creation of new liquid
bridges as discussed before in the context of the equivalent Eq.~(\ref{E3}).
We remark that here we used that the correlation $\gcHSS$
of the dry system at contact has equal contributions from positive and
negative relative velocities, immediately before and after the collision, which
is still true for the wetted elastic particles we consider. This symmetry between positive
and negative radial relative velocities is broken if one wishes to
introduce a restitution coefficient $0<\epsilon<1$ to model
inelastic collisions: the contact correlation of positive
velocities is then increased by a factor $1/\epsilon$ as compared to the negatives.

One should see clearly the very different meaning of $K$ and
$K_{\text{sites}}$. The dynamical quantity $K$ is the \emph{number
of instantaneously existing capillary bonds}:
\begin{equation}
K=2^D D \ \phi \ \frac{\gbc}{d} \ \int_0^{\scrit} \bs(s)
\ \gvol(s) \ \D s \ . \label{BridgeCoordination}
\end{equation}
$K$ rapidly decays close to zero in dilute systems. As $K$ comes
closer to the value of $K_{\text{sites}}$ in a very dense system,
the binding frequency $f_{\text{bind}}\propto \gunc \propto K-K_{\text{sites}}$
(\ref{fbind}) goes to zero because steric hindrance prohibits the formation of
further capillary contacts: $K-K_{\text{sites}}$ gives the number of
vacant sites for capillary bonds. Therefore $K_{\text{sites}}$ is the
maximum number of \emph{`docking sites'} for capillary bonds. It
is a pure geometric property 
and grows with $\scrit$, because $\scrit>0$ still allows
for a slight rearrangement of particles in the formation of new
capillary bridges without breaking existing ones. In the limit
$\scrit \rightarrow 0$, $K_{\text{sites}}$ is the number
of `contact sites'. We therefore expect $K_{\text{sites}}$ to
equal the number of exact contacts, $2D=4$. So let us compute
$K_{\text{sites}}(\phi,\scrit)$ in the following
paragraph.

\begin{figure} \begin{center}
(A) \scalebox{0.9}{\epsffile{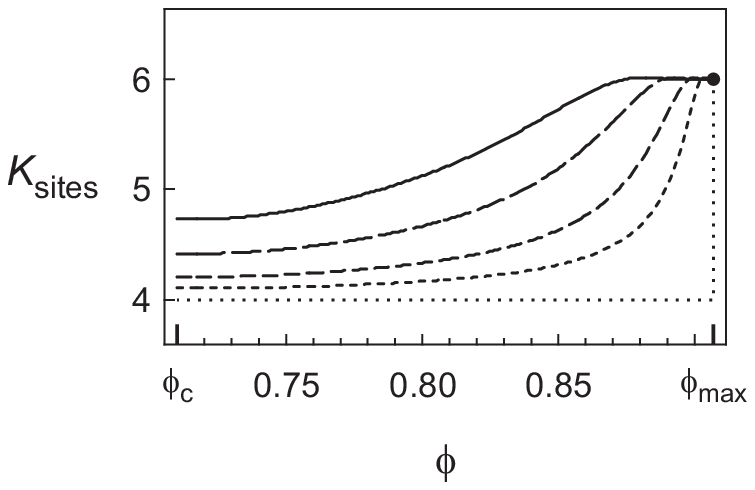}}
\qquad (B) \scalebox{0.6}{\epsffile{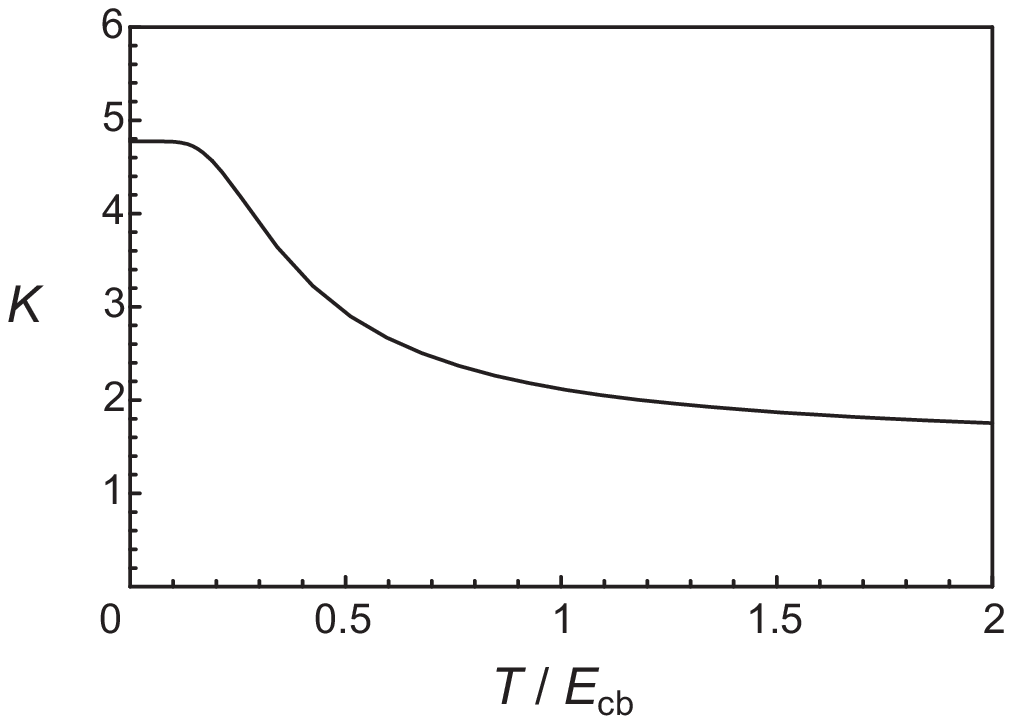}}
\caption{(A) The maximum number of
'docking sites' for capillary bonds on a particle,
$K_{\text{sites}}$, which is possible in two dimensions given the
pinch-off length $\scrit$ of capillary bridges and the
density $\phi$. $K_{\text{sites}}$ is independent
of the temperature because it is a pure geometric quantity: the number
of possible neighboring sites. Here $K_{\text{sites}}$ is shown as
a function of $\phi$ for different $\scrit$ ranging from
$\scrit/d=0.07 \text{ (solid curve), } 0.04, 0.02 \text{, to
} 0.01 \text{ (short dashes)}$. As is shown in the text,
$K_{\text{sites}}$ is the coordination number in the zero
temperature limit. While the mean coordination number $K$ rapidly
goes to zero with density for finite temperature, the zero-temperature
limit is $\geq 4$ for all densities, because the
system clusters. The dotted curve represents the limit
$\scrit \rightarrow 0$.
The longer $\scrit$ the closer $K_{\text{sites}}$ comes to six,
the number of next neighbors in two dimensions. In the limit
$\scrit \rightarrow 0$ the coordination $K_{\text{sites}}$ converges
to the number of exact contacts which is precisely four.
(B) The capillary bridge coordination $K$ drops down in the vicinity of the critical temperature
as clustered structures break up.
For this plot the mean density is chosen to be $\phi=0.75$.} \label{chpEqStKsites}
\end{center} \end{figure}

The maximum number of possible bonds, $K_{\text{sites}}$, is an
athermal function of density $\phi$ and the critical liquid bridge
length $\scrit$. We determine $K_{\text{sites}}$ from the
obvious fact, that the granular dynamics is unaffected by the
introduction of forceless bridges: for $\bs=\us$ we recover the
dry contact correlation $\gbc+\guc = \gcHSS$. This is the athermal
limit, or high temperature limit of wet granular matter.
\begin{eqnarray}
\text{High temperature limit}
\left\{ \begin{array}{l}
\bs = \us \\
\gbc+\guc = \gcHSS
\end{array} \right . \label{HTLofWGM}
\end{eqnarray}

From the hysteretic bridge system (\ref{HystereticSystem}) follows
in this forceless or high granular temperature limit
(\ref{HTLofWGM}):
\begin{eqnarray}
\left({1-
\frac{\us \gvol}{1+\gamma_{\text{pass}}}} \right) \
K_{\text{sites}} &=&
\left({1+\us \gvol}\right) \ K \text{ and} \label{KZwischenFormel}  \\
\gbc &=& \frac{\gcHSS}{1+\us \ \gvol} \ .
\label{Elimgbc}
\end{eqnarray}
The $\gamma$-functions with the argument $s$
suppressed are understood to be evaluated at $s=\scrit$.
Inserting the Eqs.~(\ref{KZwischenFormel}) and (\ref{Elimgbc}) in
(\ref{BridgeCoordination}) yields
\begin{eqnarray}
K_{\text{sites}} &=& 2^D D \ \phi \ \gcHSS
\frac{\int_{0}^{\scrit} \ \us(\scrit) \ \gvol(\scrit) \ \D s
/ d}{1-\us \gvol/{(1+\gamma_{\text{pass}})}} \label{ExactContactLimit} \\
&=& 4 \ \frac{1-\us + {\cal
O}\left(\scrit^3\right)}{{1-\us + {\cal
O}\left(\scrit^2\right)}} = 4 \ + {\cal
O}\left(\scrit\right) \ . \label{ExactContactLimitD=2}
\end{eqnarray}
In the last line we have set $D=2$, so that we could use
(\ref{ShortRangeCorr}) and (\ref{ShortRangeCorrB}). The result
(\ref{ExactContactLimitD=2}) is the second important consistency
test. Finding the number of exact contacts in the jamming limit to
equal four in (\ref{IsostaticResult}) showed the consistency of
the free-volume argument applied there. Here in
(\ref{ExactContactLimitD=2}) we find for any density that the
different function $K_{\text{sites}}$ for the number of possible
bridges sites equals four as well when $\scrit=0$. This is as
intuitively expected and a confirmation of the consistency between
the hysteretic system (\ref{HystereticSystem}) and the
near-contact pair correlation. In view of the numerical finding
$Z_{\text{sim}}=2$ for the derivative at contact of the pair
correlation (as defined in (\ref{DefinitionZ})) we remark that the
entirely analytic description by the hysteretic system gives in
general $K_{\text{sites}}=8 / Z + {\cal O}\left(\scrit\right)$,
which is why the consistency is non-trivial and the finding
$Z_{\text{sim}}=2$ fits favorably into the entire picture.


Thus the hysteretic system (\ref{HystereticSystem})
provides a direct connection between the {static} granular
properties captured in $K_{\text{sites}}$ and the granular system
in motion at positive granular temperature which we are treating in general.
We remind that $K_{\text{sites}}$ is determined by the steric self-hindrance and therefore a pure geometric property
independent of the
granular temperature. When inspecting a snapshot of a close
granular packing we can find local cases of contact coordination
($\scrit=0$) higher than four. These are fluctuations
within the granular ensemble, while $K_{\text{sites}}$ and $K$ are
mean-field quantities. Of course, for a finite bridge length
$\scrit>0$, a mean bridge coordination
$K\in(0,K_{\text{sites}})$ with $K_{\text{sites}}$ higher than
four is possible due to elongated bridges, as described by (\ref{ExactContactLimit}).
Before we evaluated the expression (\ref{ExactContactLimit}) of $K_{\text{sites}}$ for
positive $\scrit$ (plotted in Fig.~\ref{chpEqStConfirmKsites}), it
is enlightening to switch on the capillary forces in the following
section because this allows us to apply $K_{\text{sites}}$ to 'frozen' wet granular matter.


\subsection{Switching On the Force of Capillary Bridges} \label{BridgesWithForce}
Under the attraction of a liquid bridge, the pair correlation
$g^{\text{b}}(s)$ of connected neighbors falls off faster than
$g^{\text{u}}(s)$ for unbound particles, depending on the granular
temperature $T/\Ecb$ compared to the bridge energy.
The logarithmic derivative of the radial pair correlation is to be
interpreted as the effective radial force
\cite{Henderson2003,Henderson2004}, $\beta F = \pd_s \ln g(s) $,
as discussed before in section \ref{SecDeplForce}. This exponential
dependence can be justified as the solution of the
Fokker-Planck equation derived in \cite{Piasecki}. Moreover,
in the context of the hysteretic interaction of wet granular matter this
exponential factor has been successfully applied in the
case $D=1$ (cf.~Eq.~(3) in \cite{Fingerle2006}). Therefore we
proceed by switching on the liquid bridge force to the physical
value of the Minimal Capillary model \cite{Herm05},
$F_{\text{b}}=\Ecb/\scrit$, including this
exponential
in the short-range dependence of the pair correlation function for bridges neighbors:
\begin{eqnarray}
\bs(s,T) = \us(s) \ \exp{\left(-\frac{\Ecb}{T} \
\frac{s}{\scrit}\right)} \ . \label{BondStretch}
\end{eqnarray}
At low granular temperatures this exponential gives rise to
shorter average bridge lengths, and describes the reduced
probability that a bridge reaches its critical length $\scrit$. Therefore
the hysteretic system (\ref{HystereticSystem}) describes the
sticking of particles and the onset of clustering.

We have discussed in the previous section \ref{ForcelessBridges} that
steric effects in the dynamical system limit the mean number of
bonds to a maximum of $K_{\text{sites}}$, and we derived that
$K_{\text{sites}}$ converges to the number of isostatic contacts
in the limit $\scrit \rightarrow 0$.
Here this connection is put on firm grounds with a clear physical
interpretation attributed to $K_{\text{sites}}$: $K_{\text{sites}}$
\emph{is the bridge coordination $K$ of solid wet granular matter.}

\subparagraph{Proof of $K \rightarrow K_{\text{sites}}$ in the low temperature
limit.} \label{LTLproof}
Solving (\ref{HystereticSystem}) for $K/K_{\text{sites}}$, we obtain
\begin{eqnarray}
\frac{K(T,\scrit,\phi)}{K_{\text{sites}}(\scrit,\phi)}
= \frac{1}{1+X(T)/Y(T)} \ , \label{TightBindingRatio}
\end{eqnarray}
with
\begin{eqnarray}
X(T) &=& \bs(\scrit,T) \ K_{\text{sites}} \ (\gamma_{\text{pass}}+2)
\gvol
\text{ and} \\
Y(T) &=& 8 I(T) \ \phi\gcHSS \ (\gamma_{\text{pass}}+1) \ ,
\end{eqnarray}
 where $I(T)$ stands for the integral over bond states,
\begin{eqnarray}
I(T)= \int_{s=0}^{\scrit} \ \bs \gvol \ \D
s / d= \int_{s=0}^{\scrit} \
\Exp{-\frac{\Ecb s}{T \scrit}} \ \us
\gvol \ \D s / d = \frac{T
\scrit}{\Ecb d} + {{\cal O}\left(T^2\right)} \ , \label{LowTempExpansion}
\end{eqnarray}
which goes linearly to zero, while $\bs(T)\propto
\Exp{-\Ecb/T}$ vanishes for $T \rightarrow 0$ faster than
any power of $T$. Hence $X/Y \rightarrow 0$ so that
Eq.~(\ref{TightBindingRatio}) implies
\begin{eqnarray}
\lim_{T\rightarrow 0} K = K_{\text{sites}}
\end{eqnarray}
as conjectured.



This low temperature limit ($T\ll \Ecb$) is of general interest
since it represents a sticky gas of ideal spheres, which serves as
a model for aggregation in various areas of physics
\cite{PhysRevE.58.R2733} and astrophysics
\cite{PhysRevLett.85.2426}: once two particles had contact, the
remaining degree of freedom is tangential motion. The analytic
prediction of formula~(\ref{ExactContactLimitD=2}) is
$K_{\text{sites}}=4$ in the limit of exact contacts, $\scrit=0$.
In order to evaluate (\ref{ExactContactLimit}) for positive
$\scrit$ we insert the near-contact decay $\us$ given by the
general results (\ref{ShortRangeCorr}), (\ref{ShortRangeCorrB}),
and (\ref{ShortRangeCorrDepl}), setting $D=2$. The explicit
expression for $\us$ which we use throughout this article for
results without free parameters is given in the
appendix~\ref{AppendixExplicitPairCorrelation}. Here we take into
account known formulas for the contact value $\gcHSS$ at low
densities, as well as higher corrections to the free volume
theory. Inserting this expression
in (\ref{ExactContactLimit}) results in the curve shown in Fig.~\ref{chpEqStConfirmKsites}.
We have performed simulations in this low-temperature limit. The wet granular matter
was initially prepared in a gas state with $T=50\Ecb$ and cooled by the
formation and rupture of bonds. The insets in Fig.~\ref{chpEqStConfirmKsites} show final
states when the granular temperature $T$ is more than one order
of magnitude below $\Ecb$ and no further change in the configuration
was observed on exponential time scales.
The symbols in Fig.~\ref{chpEqStConfirmKsites} have been measured in this final state.
In perfect agreement with the prediction of Eq.~(\ref{ExactContactLimitD=2}),
we find in the contact limit,
$\scrit\rightarrow 0$, the coordination to be exactly $4$.
Moreover, the increase in the number of bonds per particle with the increase
of the maximal bridge length $\scrit$
is found to be in very good agreement with the simulations.

Further analytic results for high densities
are shown in Fig.~\ref{chpEqStKsites}~(A).
As is intuitively clear and shown by the family of curves in
Fig.~\ref{chpEqStKsites}~(A), the convergence of the
limit $\scrit\rightarrow 0$ is
not uniform with respect to density, since $K_{\text{sites}}$ is
pinned to the kissing number 6 of the monodisperse crystal density at $\phi_{\text{max}}$.

\section{The Equation of State}

We are now in the position to derive the equation of state,
$P=P(T,\phi)$, for wet granular matter with capillary bonds
tensile up to the rupture length $\scrit$. The cohesion of
capillary bridges will reduce the pressure as compared to a dry
hard-sphere system of equal temperature. By virtue of
Eq.~(\ref{TightBindingRatio}), we have the bridge coordination
number $K$ as a function of density $\phi$ and granular
temperature $T$. Since in the Minimal Capillary model
\cite{Herm05} the bridge force is assumed to be independent of the
bridge length $s$, the knowledge of the mean number of bridges $K$
will allow us to evaluate the reduction of the pressure due to
cohesion. Furthermore, the particle-particle collisions are
enhanced by the bridge attraction, increasing the contact
correlation. The contact correlation $\gcwet$ for wet granular
matter derives from the Eqs.~(\ref{HystereticSystem}) and
(\ref{BridgeCoordination}):
\begin{eqnarray}
\gcwet=\gcHSS \frac{(1+\gamma_{\text{pass}})(1+\bs \gvol) \ I^{\text{u}}}
{(1+\gamma_{\text{pass}}-\us \gvol)I^{\text{b}} +(2+\gamma_{\text{pass}})\bs \gvol I^{\text{u}}}
\label{ExpliciteCC}
\end{eqnarray}
with the integrals
\begin{eqnarray}
I^{\text{u/b}} = \int_0^{\scrit} \ \ubs(s) \ \gvol(s) \ \D s \ .
\end{eqnarray}
The analytic expression~(\ref{ExpliciteCC}) for the contact
correlation of wet granular matter, $\gcwet$, is indeed strictly
greater than the one of the dry system, $\gcHSS$, to which it
converges in the high temperature limit when the capillary energy
$\Ecb$ is small compared to the granular temperature $T$. This
limit follows obviously from (\ref{ExpliciteCC}) because the
functions with superscript index 'b' turn into those with 'u' for
$T\gg \Ecb$. In the low temperature limit, liquid bonds oscillate
with an amplitude proportional to the kinetic energy which equals
$T$ on average, so that the probability to find the particles at
contact, $\gcwet$, grows proportional to $1/T$, as can be derived
easily from (\ref{ExpliciteCC}) using the expansion
(\ref{LowTempExpansion}).


\subsection{Frozen Degrees of Freedom}
As the system starts to cluster at temperatures close to $\Ecb$,
voids remain between the clusters
with linear dimensions large compared to the particle diameter.
Clearly, this growing length scale, which is set by the sizes of clusters
and voids, is not captured by the short-range behavior of the pair
correlation function. Here we advance the theory beyond the level of
two-particle correlations to take correlation on large scales,
such as the collective particle motion in a cluster, in an approximative
fashion into account.

\begin{figure} \begin{center}
\epsffile{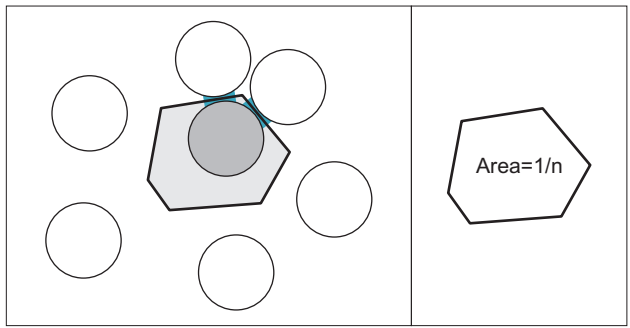} \caption{A local
configuration of two-dimensional wet granular
matter at moderate density. Since there is one
particle in each Vorono\"i cell, the mean area equals the inverse
density. In section \ref{DryDense} we have used the Vorono\"i
tessellation to compute the derivative of the pair correlation at
contact for a dry and dense system. For such a dense system, the
Vorono\"i cell resembles a hexagon with a size proportional to
$(d+s)^2$, where $s$ is the particle separation. The cell borders
are at half surface separation for polydisperse diameters
(not half center distance), so that each cell contains one particle
completely.}
\label{chpEqStGraphic3}
\end{center} \end{figure}

The collective motion of a cluster is due to stable capillary bonds
which impose constraints, such that the internal degrees of freedom of clusters
are frozen. Since $K$ is the number of instantaneous capillary bridges
of which the fraction $\text{erf}\left(\sqrt{{\Ecb}/{T}}\right)$ with kinetic
energies below $\Ecb$ forms stable bonds, we have
\begin{eqnarray}
K_{\text{frozen}} &=& K \
\text{erf}\left(\sqrt{\frac{\Ecb}{T}}\right) \label{Kfrozen}
\end{eqnarray}
for the number of frozen degrees of freedom.

We are interested in the density of the remaining degrees of freedom.
The idea is simple and powerful:
As a general mathematical property of triangulations, there are on
average precisely six Vorono\"i neighbors \cite{Meijering1953},
independent of density or ordering.
In Fig.~\ref{chpEqStGraphic3} we can observe that the Vorono\"i
neighbors with stable bonds contribute less to the area $1/n$ of
the Vorono\"i cell. This picture suggest a two-fluid model
with frozen and free neighborhoods as the two constituents.
The fraction of frozen and free triangulation bonds is proportional to
$K_{\text{frozen}}$ and $K_{\text{free}}$ respectively, and
the area contributions associated to each bond sum up to the
total size of the Vorono\"i cell:
\begin{eqnarray}
K_{\text{frozen}} + {K_{\text{free}}} &=& {6} \label{DualFluidCond1} \\
\frac{K_{\text{frozen}}}{n_{\text{frozen}}} + \frac{K_{\text{free}}}{n_{\text{free}}} &=& \frac{6}{n} \ . \label{DualFluidCond2}
\end{eqnarray}
Because of this reciprocal sum rule for the densities one may call this a reciprocal two-fluid model.

The density $n_{\text{frozen}}$ of the stable bond component follows
analogously to (\ref{DensityCondition}) when
averaged with the additional exponential factor
(\ref{BondStretch}) due to the capillary force:
\begin{eqnarray}
\left< \left(1+\frac{s}{d} \right)^D \right>_{\text{frozen}} &=& \frac{n_{\text{J}}}{n_{\text{frozen}}} \label{BondDensityCondition} \\
\left< \dots \right>_{\text{frozen}} &=& \frac{\int_0^{\scrit} \ \dots \ \gamma_{\text{frozen}}(s) \ \gvol(s) \ \D s}{\int_0^{\scrit} \ \gamma_{\text{frozen}}(s) \ \gvol(s) \ \D s} \\
\gamma_{\text{frozen}}(s) &=& \exp{\left(-\phi \ \gcHSS \left[\left(1+\frac{s}{d} \right)^D-1\right] -\frac{\Ecb}{T} \
\frac{s}{\scrit}\right)} \label{gammaFrozen}
\end{eqnarray}
Without affecting the leading order in $s/d$ one is free to
replace the last $s$ in the exponent (\ref{gammaFrozen}) by
$s+s^2/(2d)$, so that the integral (\ref{BondDensityCondition}) is
elementary resulting in
\begin{eqnarray}
&&\frac{n_{\text{J}}}{n_{\text{frozen}}}-1 = \left[\left(1+\frac{\scrit}{d} \right)^D-1\right] \ \left(\frac{1}{\alpha}-\frac{1}{\Exp{\alpha}-1}\right) \label{nfrozen} \\
&&\text{with } \alpha=\left[\left(1+\frac{\scrit}{d} \right)^D-1\right] \left(\phi \ \gcHSS + \frac{\Ecb}{T} \frac{d}{D \scrit}\right) \label{AlphaExponentBelDim}
\end{eqnarray}
We point out that Eq.~(\ref{nfrozen}) implies Eq.~(4) in \cite{Fingerle2006} for $D=1$.

From the Eqs.~(\ref{TightBindingRatio}), (\ref{Kfrozen})-(\ref{DualFluidCond2}),
and (\ref{nfrozen})-(\ref{AlphaExponentBelDim}) follows the density of degrees
of freedom which are not frozen out by capillary bonds,
$n_{\text{free}}(T,\scrit,\phi)$.
One may regard $n_{\text{free}}$ as the density of clusters.


We remark that the two-fluid model of neighborhoods is the only
concept presented in this theory of wet granular matter which
cannot be generalized in a straight forward manner to three
dimensions, because for $D=3$ the number of Vorono\"i neighbors
(double counted per particle) is not a universal constant (such as
$6$ for $D=2$ and $2$ for $D=1$), but depends on the granular
order (reaching its minimum value $12$ for close packing and its
maximum of approximately $15.5$ in the ideal gas limit)
\cite{Aste2004}. The reason for this is that three-dimensional
space cannot be filled with tetrahedrons, while flat space can be
tiled by triangles. As a consequence, the number of constituents
in the two-fluid model of neighborhoods would not be conserved for
$D=3$ and the numerator on the right-hand side of
(\ref{DualFluidCond2}) is not a constant.

\subsection{The Pressure of Wet Granular Matter}

\begin{figure} \begin{center}
\scalebox{1.0}{\epsffile{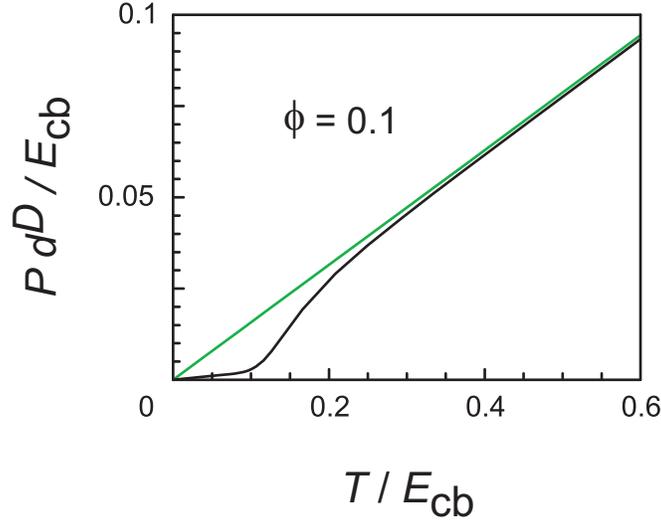}}
\caption{The pressure $P$ of wet granular matter is
shown as function of the granular temperature $T$.
The dimensionality is $D=2$ and
the covered area fraction is $\phi=0.1$,
so that at high temperatures
the system is a dilute gas.
The maximum bridge length is $\scrit=0.07 d$.
The behavior below the critical
temperature $T_{\text{c}}=0.274 \Ecb$ of wet granular can be
understood in the following way: the system agglutinates to clusters.
With these effective particles the pressure is reduced according
to the reduced number density of effective particles. The breakup of clusters
is reflected by the rising pressure around $T_{\text{c}}$.
The straight line is the athermal pressure of hard discs,
$P^{\text{dry}}=n \gcwall T$ which is reached asymptotically when the granular
temperature is higher than the energy scale $\Ecb$ set by the capillary interaction.
} \label{chpEqStGraphic10a}
\end{center} \end{figure}

\begin{figure} \begin{center}
\scalebox{1.0}{\epsffile{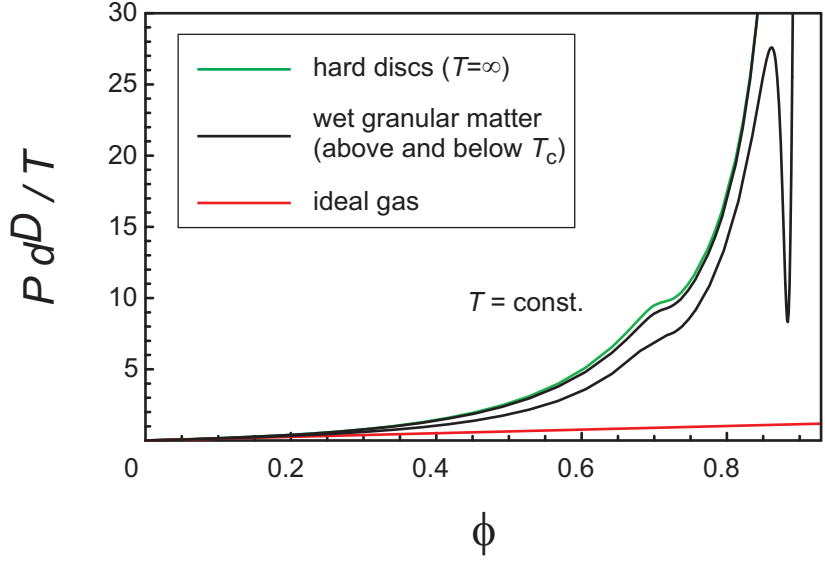}}
\caption{Isotherms of wet granular matter for the realistic
rupture length $\scrit=0.07 d$. In the high temperature limit the liquid bridges
forfeit their influence on the dynamics, so that the equation of
state reduces to the hard sphere pressure. This can be seen by the
two black isotherms of wet granular matter, of which the higher is
at $T=\Ecb$ and converges to the green curve in the limit $T\gg
\Ecb$. The lower black isotherm is at $T=0.2 \Ecb$ and exhibits an
unstable branch. The critical point is at $T_{\text{c}} \approx
0.274 \Ecb$ (cf.~Fig.~\ref{chpEqStGraphicSchleife} for a close-up).} \label{chpEqStGraphic10}
\end{center} \end{figure}

\begin{figure} \begin{center}
\scalebox{0.6}{\epsffile{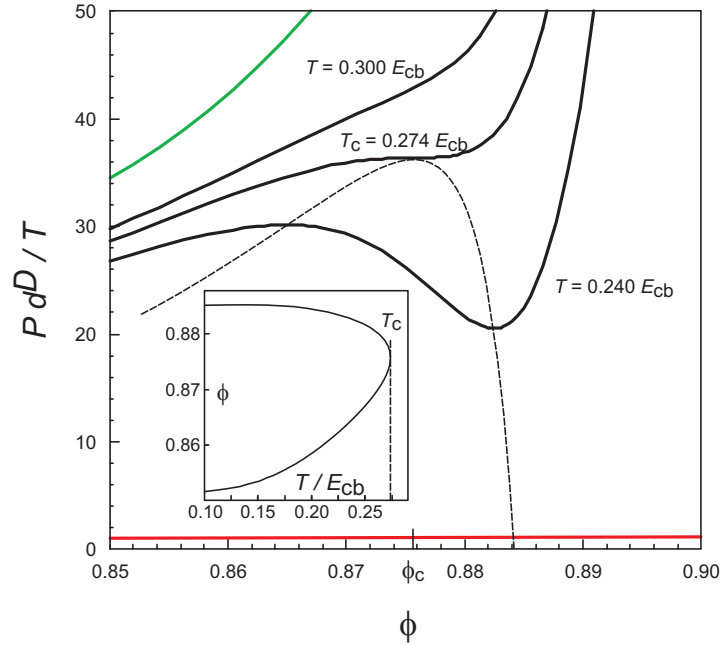}}
\caption{A close-up of the transition region in wet granular matter.
The dashed line in the main panel is the spinodal of
the homogeneously driven wet granular
system in $D=2$ dimensions. The solid black lines are wet granular
isotherms around the critical point, which is located at
$T_{\text{crit}} =  0.273(5) \Ecb$ for $\scrit=0.07d$. The
change of the critical point with the amount of added liquid
(represented by $\scrit$) is shown in Fig.~\ref{chpEqStKritPkt}.
The curve in the upper left corner is the athermal
pressure $P^{\text{dry}}$ of the hard disc system
\cite{LudingZustandsgleichung} without liquid bridges, and the
line at the bottom is the ideal gas pressure ($P^{\text{id}}
d^D(\phi)$ has a defined slope). $P^{\text{dry}}=\gcHSS \
P^{\text{id}}$ is increased compared to the ideal gas by the
Enskog factor $\gcHSS$. The pressure of wet granular matter is
reduced compared to the dry system $P^{\text{dry}}$ due to the
capillary cohesion. The inset shows the spinodal in the temperature-density
plane, where the critical temperature can be clearly determined.}
\label{chpEqStGraphicSchleife}
\end{center} \end{figure}

\begin{figure} \begin{center}
\scalebox{1.0}{\epsffile{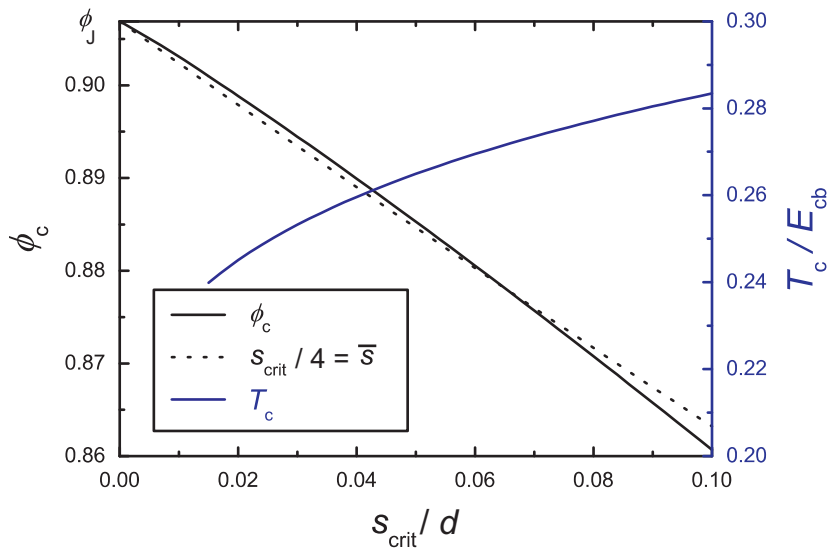}} \caption{The
influence of the rupture length $\scrit$ on the position of the
critical point in the phase diagram
Fig.~\ref{chpEqStGraphicSchleife} of wet granular matter. The
position of the critical point is described by the critical
parameters $(\phi_{\text{c}},T_{\text{c}})$, which are plotted on
the left and right vertical axis respectively. Solid lines result
from the full theory (\ref{FinalPressureExpression}) by solving
for the intersection of $\pd_\phi P(\phi,T)=0$ and $\pd^2_\phi
P\phi,T=0$. For the critical temperature we find a very mild
variation with the rupture length, so that over the entire
physically relevant range of capillary interaction we have
$T_{\text{c}}\approx \Ecb/4$. The influence of the rupture length
$\scrit$ on the critical density $\phi_{\text{c}}$ can be
understood very clearly with the help of the dashed line. The
critical density is such that the mean particle separation
$\overline{s}=d
\left(\sqrt[D]{\phi_{\text{J}}/{\phi_{\text{c}}}}-1\right)$ scales
with the rupture length $\scrit$. This shows that both intrinsic
characteristics of the capillary interaction, the rupture length
and the bridge energy $\Ecb$, determine the critical point of wet
granular matter.} \label{chpEqStKritPkt}
\end{center} \end{figure}


Here we arrive at the pressure $P(T,\phi)$ using the density $n_{\text{free}}(T,\phi)$ (\ref{DualFluidCond1}) of degrees of freedom, the coordination $K(T,\phi)$ (\ref{TightBindingRatio}), and the contact correlation $\gcwet(T,\phi)$ (\ref{ExpliciteCC}).
The pressure is the trace of the stress tensor
\begin{equation}
P = - \frac{1}{D} \ \text{tr} \underline{\underline{\sigma }} \ . \label{DefinitionDruck}
\end{equation}
The stress tensor $\underline{\underline{\sigma
}}=\underline{\underline{\sigma
}}^{\text{kin}}+\underline{\underline{\sigma }}^{\text{force}}$
describes the flow of momentum. The kinetic term has components $
\sigma^{\text{kin}}_{{i,j}}=-\sum_k^N \left<m v^{(k)}_i v^{(k)}_j
\ \delta({\bf r}-{\bf r}^{(k)})\right> $. With the granular
temperature $T=\left<m v_i v_i\right>$, its trace yields $n T$ for
uncorrelated particle motion (as in an ideal gas). In general we
have the kinetic contribution
\begin{eqnarray}
P^{\text{kin}}=n_{\text{free}} T \label{ClusterKinetic}
\end{eqnarray}
wherein there frozen degrees of freedom have been taken out. For
moderate densities, one may interpret (\ref{ClusterKinetic}) as
the kinetic contribution to the pressure due to a gas of clusters.

The interparticle forces $\bf F$ give rise to the Cauchy tensor
$\underline{\underline{\sigma }}^{\text{force}}$, which is the
tensor product of the center-to-center vector $\bf r$ and the pair
force $\bf F$,
\begin{equation}
\underline{\underline{ \sigma }}^{\text{force}} =\frac{n_{\text{free}}}{2} \left<{\bf F} \otimes {\bf r} \right> \ , \label{FreeCauchyTensor}
\end{equation}
so that $\underline{\underline{ \sigma }}^{\text{force}}$ is
diagonal for radial forces. The factor $1/2$ assigns half of the
momentum current to either of the interaction particles, i.e.
${\bf r}/2$ may be seen as the transport vector within the
Vorono\"i cell. The Cauchy tensor (\ref{FreeCauchyTensor}) has
contributions only by the unfrozen pairs of particles with density
$n_{\text{free}}$, because in frozen neighborhoods the repulsive
momentum exchanged in collisions is exactly balanced by the bridge
attraction under the time average on the right-hand side of
(\ref{FreeCauchyTensor}).

A comment on the significance of the reciprocal two-fluid model as represented
by Eq.~(\ref{DualFluidCond1}) and (\ref{DualFluidCond2}) is
in order here. We consider for instance a compressed state of wet
granular matter with $K_{\text{frozen}}$ around five and
$K_{\text{free}}$ around unity. While $K_{\text{free}}$ is small,
the prefactor $n_{\text{free}}$ in (\ref{FreeCauchyTensor}) is not
necessarily small. From (\ref{DualFluidCond1}) and
(\ref{DualFluidCond2}) follows that both, $n_{\text{free}}$ and
$n_{\text{frozen}}$, converge to $n_{\text{J}}$ as the system gets
jammed ($n \rightarrow n_{\text{J}}$), so that the repulsive
dominated state is correctly described by the Cauchy tensor
(\ref{FreeCauchyTensor}) which grows beyond all bounds as $n
\rightarrow n_{\text{J}}$. If one had (in contradiction to the
additivity of areas) summed up densities linearly
instead of the reciprocal sum rule (\ref{DualFluidCond2}), the
free density would vanish or could even become negative under such
conditions.

It is finally easy to determine the time average
on the right-hand side of (\ref{FreeCauchyTensor}) for the two different forces acting in
wet granular matter,
the delta-force in collisions of hard particles and the flat force $F_{\text{cb}}=\Ecb/\scrit$
of the capillary bonds. In a collision at time $t_{\text{coll}}$ the radial momentum $\Delta {\bf p}$
is transferred instantaneously:
\begin{equation}
\left<{\bf F}_{\text{coll}} \otimes {\bf r} \right> = \left<\Delta {\bf p} \otimes {\bf r} \ \delta(t-t_{\text{coll}})\right> = \E \ \left<\Delta {p} \ ({\bf r},-{\bf v}) \ \theta\left(({\bf r},-{\bf v})\right) \ \delta(r-d) \right> = -\E \ \gcwet n \sigma_D d^{D} T \ . \label{StossVirial}
\end{equation}
In the last equality the $\delta$-function gives rise to the contact correlation
$\gcwet$ and the trivial integration of angles leaves $\sigma_D d^{D-1}$.
$\E$ is the unity matrix and $\theta$ is the Heaviside step function.
Inserting (\ref{StossVirial}) in (\ref{FreeCauchyTensor}) and taking the trace
(\ref{DefinitionDruck}) yields
\begin{equation}
P_{\text{coll}} = 2^{D-1} n_{\text{free}} \ T \ \phi \ \gcwet \ .
\end{equation}
The cohesive virial due to capillary bridges is
\begin{equation}
\left<{\bf F}_{\text{cb}} \otimes {\bf r} \right> = \left<K \frac{\Ecb}{\scrit} \ \frac{{\bf r}\otimes {\bf r} }{r}\right> = \frac{\E}{D} \ K\frac{\Ecb}{\scrit} \ \left<d+s\right> \approx \frac{\E}{D} \ K\frac{\Ecb}{\scrit} \ d \ . \label{BridgeVirial}
\end{equation}
Hence the final result
\begin{equation}
P= n_{\text{free}} T \ \left( 1+2^{D-1} \ \phi \ \gcwet \right) \ -  \ n_{\text{free}}
\Ecb \ \frac{K}{2 D} \frac{d}{\scrit} \  , \label{FinalPressureExpression}
\end{equation}
where the last term is the bridge cohesion (\ref{BridgeVirial}).
Since $n_{\text{free}}$, the contact correlation $\gcwet$ and $K$
have been derived explicitly in (\ref{DualFluidCond1}),
(\ref{ExpliciteCC}) and (\ref{TightBindingRatio}) as functions of
$\phi$ and $T$, we have the equation of state for wet granular
matter, $P=P(\phi,T)$.

The Figs.~\ref{chpEqStGraphic10a} and \ref{chpEqStGraphic10}
show the analytic result (\ref{FinalPressureExpression}) as a function
of the granular temperature $T$ and the density $\phi$. 
In the high temperature limit wet granular matter
behaves as a hard-spheres system. Below the
critical point granular clusters are predicted to segregate due to
the mechanically unstable branch of the pressure as a function of density,
which appears in Fig.~\ref{chpEqStGraphic10} below the critical temperature.
Figure~\ref{chpEqStGraphicSchleife} provides a close-up of the
critical point of wet granular matter and its spinodal. The
critical density of this transition is high, because the
particles have to be close enough in order to form a dynamical capillary
network. As we show in Fig.~\ref{chpEqStKritPkt}, the critical
density is determined by the length scale of capillary bridges, such that
the rupture length $\scrit$ scales with the mean particle
separation $\overline{s}$.
Moreover, the rupture length is approximately
four times the mean particle separation, $\scrit
\approx 4 \overline{s}$ (dashed line shown in Fig.~\ref{chpEqStKritPkt} for
comparison). This result is to be compared with the
very same ratio for the reported critical density of the
un-clustering effect \cite{Fingerle2006}: in the free cooling of dense
one-dimension wet granular matter, the granular network was found to break up
into granular droplets which precipitate out of the homogenous intial state,
as soon as the density exceeded a critical value. This critical
density was shown numerically and analytically to be set by
$\scrit \approx 3 \overline{s}$ \cite{Fingerle2006}.
The different prefactor is due to the additional
cooling dynamics and the dimensionality $D=1$. The theory of
wet granular matter presented in this work
predicts this transition to persist in higher dimensions.

As we shorten the rupture length $\scrit$ (which can be easily
done experimentally by evaporating the wetting liquid), the dry
system is approached in such a way that the spinodal narrows in
the $T-\phi$ plane and is shifted to the jamming point, where it
eventually shrinks to a line and vanishes.
Figure~\ref{chpEqStKritPkt} shows the convergence of the critical
density to the jamming density. Since the capillary bridge regime
sets an upper limit on the rupture length, the critical point is
confined on the density axis between the ordering transition at
$\phi_{\text{o}}$ and the jamming density $\phi_{\text{J}}$. The
critical temperature almost exclusively depends on the bridge
energy, according to $T_{\text{c}}\approx \Ecb/4$, over the entire
capillary regime.

With this discussion of transitions occurring in wet granular
matter the presentation of our theory for wet granular matter is
completed. The reader may find in
appendix~\ref{AppendixSelfConsistentK} a brief methodical
extension of the theory where a self-consistent equation is
derived for future works.

\section{Conclusion}
Starting with the hard-sphere fluid, an expression
(\ref{ShortRangeCorr}) for the narrowing of the near contact pair
correlation was derived, which describes in the jamming limit the
delta-peak of $2D$ isostatic contacts per particle, in agreement
with the excepted value of simulations. In the gas and fluid
regime the fall-off predicted by this expression for the pair
correlation at contact was found to be well confirmed by
simulations. We then addressed the nonequilibrium case of wet
granular matter by the introduction of capillary bridges which are
formed hysteretically. The description in terms of the
pair-correlation function was extended with six different
non-vanishing correlation coefficients which take the bridge
status into accoount and allow for the hysteretic dissipative
dynamics. The coordination number of bonds was computed
analytically as a function of the rupture length of the capillary
bridges, the granular temperature, and the density. The limiting
case of strong bonds led to the sticky gas dynamics for which
simulations have been performed which showed very good agreement
with the analytic prediction of the coordination number. Based on
the derived expressions for the contact correlation and the bridge
coordination, we finally computed the pressure of wet granular
matter analytically as a function of density and granular
temperature. Here a method was put forward, which describes the
effective degrees of freedoms in order to take the correlated
motion of particles glued to clusters into account.
The isotherms of wet granular matter were found to have an
unstable branch which gives rise to the segregation of dense
clusters. The critical temperature of this transition was derived
to be approximately one quarter of the capillary bond energy. The
critical density is directly related to the pinch-off distance of
the capillary bridges. A close relation to the un-clustering
effect reported in one dimension \cite{Fingerle2006} was shown,
for which reason this effect persists also in higher dimensions.

It will be interesting to probe the critical point of wet granular
matter experimentally and by direct simulations. As we have shown,
the position of the critical point is determined by the length and
energy of the capillary bridges. These quantities can be
controlled very accurately in an experiment of shaken wet granular
matter. Therefore the measurement of the critical temperature will
allow to discern between extensions such as the nonlinear coupling
discussed in appendix~\ref{AppendixSelfConsistentK}.

Future analytic work includes the background contribution $\gB$ in
the dense regime, since our numerics indicate that the pair
correlation is flatter near the contact as predicted by $\gA$
alone. This task might be addressed in conjunction with the
analogous background contribution in three dimensions, for which
in the jamming limit an integrable power-law divergence,
$\gB\propto 1/s^{\delta}$, has been reported in numerical studies
(with $\delta=0.5$ \cite{Silbert2002} or $\delta=0.6$
\cite{Donev2005}) and experiments \cite{Aste2005b}, but is as well
lacking a theoretical explanation at present.

\paragraph*{Acknowledgments}
Discussions with Martin Brinkmann, Svenja Hager,
J\"urgen Vollmer, Klaus R\"oller and Mario Scheel
are greatfully acknowledged.

\appendix



\section{The Background Contribution $g_{\text{B}}$} \label{AppendixgB2D}
\subsection{The Weighting Factors}
With $g_{\text{A}}$ in (\ref{ShortRangeCorr}) we considered the four
(cf. Eq.~\ref{IsostaticResult}) A-neighbors, which form isostatic
contacts at jamming, $\sA\rightarrow 0$ for $\phi\rightarrow
\phi_{\text{J}}$ (\ref{JammingOfA}), and are separated by $\sA$
according to Eq.~(\ref{FreeVolumeTheory1}) before jamming.
Analogously, the separation $\sB$ of the two B-neighbors is
weighted by
\begin{eqnarray}
P_{\text{B}}(\sB) \propto \exp \left( -\frac{\left(1+\frac{\sB}{d}
\right)^2 -1 }{{\phi_{\text{max}}} / {\phi} - 1} \right) \ .
\label{ExpoAreaFactorForB}
\end{eqnarray}
While in Eq.~(\ref{FreeVolumeTheory1}) the denominator in the
exponential is $c_{\text{A}} = \phi_{\text{J}}/\phi-1$ so that
$\sA\rightarrow 0$ at the jamming density, in
Eq.~(\ref{ExpoAreaFactorForB}) the denominator is $c_{\text{B}} =
\phi_{\text{max}}/\phi-1$ since the blocked B is only forced to
form a contact, $\sB\rightarrow 0$, for a perfect crystal with $\phi \rightarrow
\phi_{\text{max}}$. Of course this limit is kinematically
unreachable because the system comes to rest at the jamming
density $\phi_{\text{J}}<\phi_{\text{max}}$.
$\phi_{\text{max}}$ would be reached.
We note that $c_{\text{B}} $ is a small dimensionless quantity: for
$\phi>\phi_{\text{o}}=0.71$ we have
$0<c_{\text{B}}<0.2774$.

Close to jamming, the B-neighbors are fixed in space by particles
other than the reference particle. Except for arch-like
constructions which are rare for frictionless particles, and would
include second Vorono\"i neighbors keeping B at a separation
larger than our region of interest, $\sB > \scrit$, this
hindrance is due to the A-neighbors. Therefore the probability
$g_{\text{B}}(\sB)$ to find a B-neighbor at separation $\sB$ from
the reference particle (sketched with hatching in
Fig.~\ref{ConfigSpaceOfABA}) is given by the integral over all configurations
where four A-neighbors hinder two B-neighbors.

\begin{figure} \begin{center}
\scalebox{0.7}{\epsffile{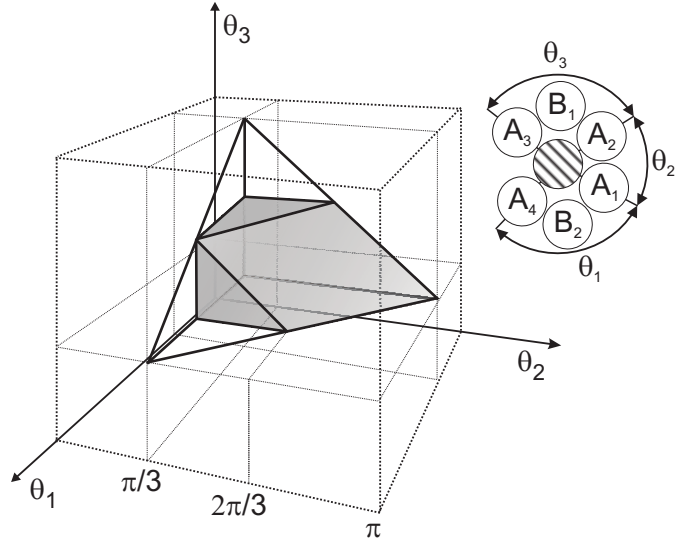}} \caption{The angular
configuration space of four neighbors close to the reference
particle. These we denote as A-neighbors. The faceted inner subset
shown in gray is the subspace conditioned to the property that two
further particles, the B-neighbors, are hinder by the
A-particles in approaching the reference particle. The projection
of this subset onto an $\theta$-axis (for the angle between a
blocking A-pair, $\theta_1$ or $\theta_3$ in this example) gives
rise to a linear configuration space factor $C(\theta)$. Obviously a
B-neighbor acts like a wedge driven between two A-neighbors, and
therefore increases $\theta$. This is taken into account by the
weighting factor $P_{\text{B}}(\sB)$ which favors shorter separations $\sB$
between the particle B and the reference particle, depending on
the density $\phi$.} \label{AngleDistribution}
\end{center} \end{figure}

The configurations will be weighted by a phase space factor $C$
and the exponential factor $P_{\text{B}}$. We are above the
ordering density $\phi_{\text{o}}$, so that the neighborhood has
(by definition of the phase) hexagonal order as sketched in the
inset of Fig.~\ref{AngleDistribution}. Projecting the
configurations with the two B-neighbors blocked (gray subset in
Fig.~\ref{AngleDistribution}) on a single $\theta$-axis, we find
the configuration space factor
\begin{eqnarray}
C(\theta)=\frac{3 (5 \pi -6 \theta )}{2 \pi ^2} \ .
\label{ConfFactor}
\end{eqnarray}
In the sequel we abbreviate
\begin{eqnarray}
\gvol(s) \propto 1+s/d \ . \label{VolFactor}
\end{eqnarray}
for the volume factor (\ref{ExpoAreaFactorForB}) in $D=2$. Wide gaps of length $\sB$ are exponentially suppressed by $P_{\text{B}}$.

\subsection{The Configuration Space}
\begin{figure} \begin{center}
\scalebox{0.7}{\epsffile{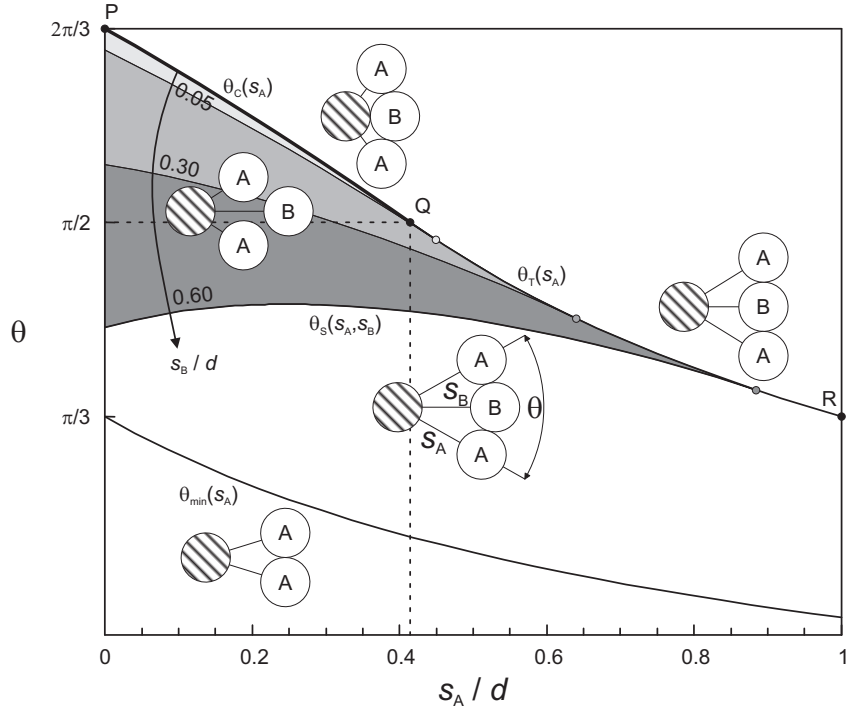}} \caption{A section of the
configuration space of neighboring particles. Within the gray
domain the particle denoted by B is blocked: the two neighbors
labelled A sterically hinder the particle B from approaching the
reference particle (shaded). Only at the boundary
$\theta_{\text{C}}(\sA)$ (curve $\overline{\text{PQ}}$ ranging
from $[\sA,\theta]_{\text{P}}=[0,2\pi/3]$ to
$[\sA,\theta]_{\text{Q}}=[(\sqrt{2}-1)d,\pi/2]$) the B-neighbor
can touch the reference particle. The probability
$g_{\text{B}}(\sB)$ to find a B-neighbor at a separation $\sB$
follows from integrating over the gray domain, which grows with
increasing $\sB$. The lower bound, $\theta_{\text{S}}(\sA,\sB)$,
is plotted for the values $\sB=0.05, \ 0.30, \ \text{and} \ 0.60$.
Large areas spanned by this neighborhood are exponentially rare
the higher the mean density $\phi$, so that the probability
distribution in this plot concentrates in the vicinity of the
upper left corner P as we come closer to the jamming limit. At the
line $\overline{\text{QR}}$ the B-neighbor slips through and turns
into an A-neighbor, so that $\overline{\text{QR}}$ is the transit
to another jamming island in configuration space. The
corresponding transition rate is proportional to the probability
density along $\overline{\text{QR}}$ and therefore vanishes in the
jamming limit.} \label{ConfigSpaceOfABA}
\end{center} \end{figure}

Let us now address the configuration space plotted in
Fig.~\ref{ConfigSpaceOfABA}. If the opening angle $\theta$ of the
A-neighbors exceeded $\theta_{\text{T}}(\sA)$,
\begin{eqnarray}
\cos \frac{\theta_{\text{T}}(\sA)}{2} &=& \frac{\sqrt{\sA (2d +
\sA)}}{d+\sA} \ , \label{thetacontact}
\end{eqnarray}
the B-particle could slip through and turn into an A-neighbor,
which is defined by having a free path towards the reference
particle. This transition corresponds to the neck connecting
different jamming island in the configuration space. Only along
the line ($\overline{\text{PQ}}$ in Fig.~\ref{ConfigSpaceOfABA})
defined by $\theta=\theta_{\text{C}}(\sA)$,
\begin{eqnarray}
\cos \frac{\theta_{\text{C}}(\sA)}{2} &=& \frac{\sA+d}{2d}
\label{thetatransition} \ ,
\end{eqnarray}
the B-neighbor can touch the reference particle, so that $\sB=0$.
The Eqs.~(\ref{thetacontact}) and (\ref{thetatransition}) define
the upper boundary of the domain of integration for all $\sA$,
\begin{eqnarray}
\theta_{\text{max}}(\sA)=
 \left \{\begin{array}{ll}
\theta_{\text{C}}(\sA) & \quad \sA/d \leq \sqrt{2}-1 \\
\theta_{\text{T}}(\sA) & \quad \sA/d \geq \sqrt{2}-1 \\
 \end{array} \right. \ , \label{IntBound1}
\end{eqnarray}
which is continuously differentiable but not smooth at the point
Q.

The lower boundary is
\begin{eqnarray}
\cos \frac{\theta_{\text{S}}(\sA,\sB)}{2}
&=&\frac{(\sA+d)^2+\sB^2+2d \sB}{2 (\sA+d) (\sB+d)} \ ,
\label{thetasteric}
\end{eqnarray}
where B hits A.

\begin{figure} \begin{center}
\scalebox{.3}{\epsffile{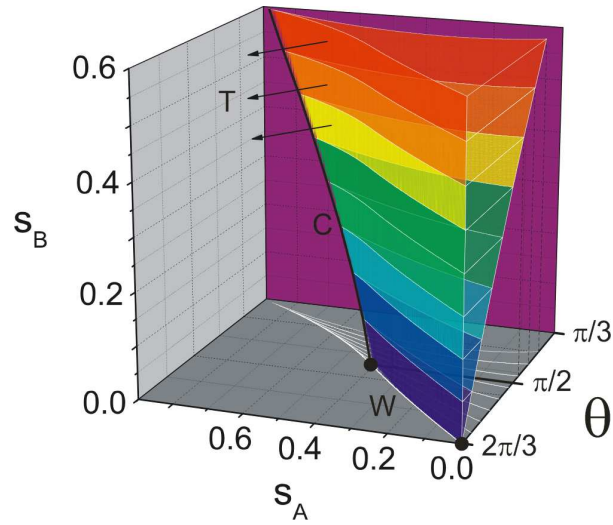}}
\caption{The $\sA$-$\theta$-plot of Fig.~\ref{ConfigSpaceOfABA}
with the full $\sB$ dependence shown on the additional vertical
axis.} \label{IntegrationDomain3D}
\end{center} \end{figure}

The simple lower bound on $\theta$,
\begin{eqnarray}
\cos \frac{\theta_{\text{min}}(\sA)}{2} &=&
\sqrt{1-\left(\frac{d/2}{d+\sA}\right)^2} \ ,
\end{eqnarray}
which ensures that the A-neighbors do not overlap is without
applicatory relevance, as it implies that the B-neighbor is pushed
out to $\sB/d>\sqrt{3}-1\approx0.73$. This is suppressed in the
dense regime $\phi>\phi_{\text{o}}$ by the factor $F$ of
Eq.~(\ref{ExpoAreaFactorForB}).

The configuration space ends to its right in a cusp where the
lower and upper bound intersect at
\begin{eqnarray}
\sA^{\text{cusp}}(\sB)=\sqrt{\sB^2+2 d \sB+2d^2}-d \ . \label{IntBound3}
\end{eqnarray}
This cusp converges to the point Q for $\sB\rightarrow 0$.

\begin{figure} \begin{center}
(A) \scalebox{0.9}{\epsffile{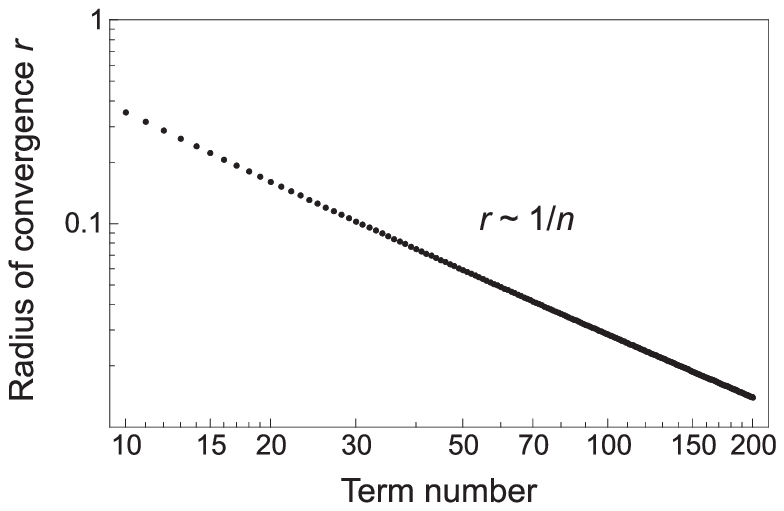}} \qquad (B)
\scalebox{0.9}{\epsffile{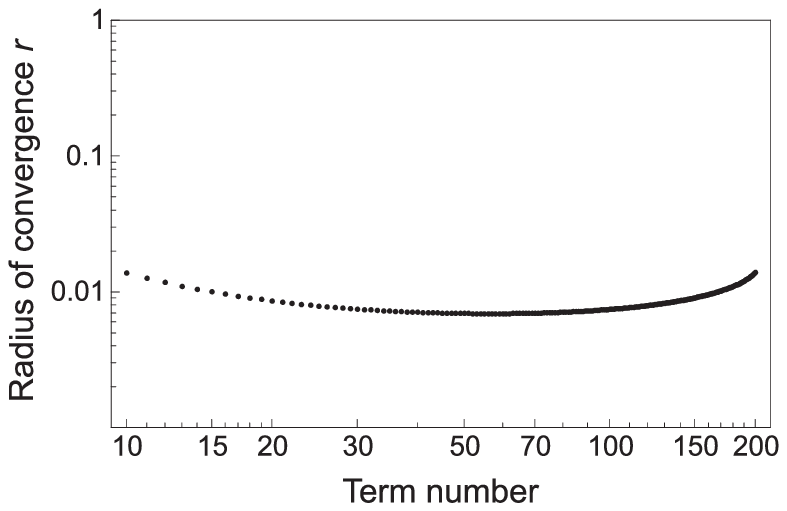}} \caption{The
radii of convergence $r_k$ for expansions around the jamming
point. The contribution blocked B-neighbors give to the pair
correlation can be expanded in a series around the jamming point,
$c_{\text{A}}=0$. The radius of convergence is given by the
Cauchy-Hadamard formula $r_j=1/\sqrt[j]{K_j}$ for the term $K_j
(\sB/d)^j$. (A) The asymptotic divergence of the $R$-series in
(\ref{FormOfIR}) poses no practical problem since few terms (less
than 10) give sufficient accuracy. (B) The $S$-series in
(\ref{FormOfIR}) converges.} \label{ConvRadius}
\end{center} \end{figure}

With the integration bounds (\ref{IntBound1}), (\ref{thetasteric}), (\ref{IntBound3}), and the weighting factors
(\ref{ExpoAreaFactorForB}), (\ref{ConfFactor}) we have
\begin{eqnarray}
g_{\text{B}}(\sB) &=& {\cal N} \ P_{\text{B}}(\sB) \left[ \int_0^{\sA^{\text{cusp}}(\sB)}
\D \sA \ P_{\text{A}}(\sA) \; \gvol(\sA) \
\int^{\theta_{\text{max}}(\sA)}_{\theta_{\text{S}}(\sA,\sB)} \D
\theta \ C(\theta) \right]^2 \label{BlockingConfigIntegral} \\
&=& {\cal N} \ P_{\text{B}}(\sB) \left[ \frac{\sB}{d} I_1(n) +
\left(\frac{\sB}{d}\right)^2 I_2(n) + {\cal
O}\left(\left(\frac{\sB}{d}\right)^3\right) \right]^2 \ .
\end{eqnarray}
We emphasize that the configuration space $(\sA,\theta)$ describes
the relative position of one A-neighbor sketched symmetrically in
Fig.~\ref{ConfigSpaceOfABA}. Since there are two independent
A-neighbors involved, their configuration is the direct product
$({\sA}_1,\theta_1)\times({\sA}_2,\theta_2)$. On this account the
configuration integral is squared in
(\ref{BlockingConfigIntegral}), with the important consequence
that the leading order in $g_{\text{B}}(\sB)$ is quadratical. The
normalization constant ${\cal N}$ is determined by the knowledge
that there are two B-neighbors. While the exponential prefactor
dominates the long range decay, we expand the near-contact
increase in $\sB/d$. Substituting the dimensionless area
$z_{\text{A}}=\left((1+\sA /d)^2-1\right)/c_{\text{A}}$ for
integration in favor of the particle separation $\sA$, the
expressions $I_i$, $i=1,2$ are of the form
\begin{eqnarray}
I_i = c_{\text{A}} \int_0^{1 / c_{\text{A}} } \ \Exp{-z_{\text{A}}} f_i(c_{\text{A}}
z_{\text{A}}) \ \D z_{\text{A}}
\end{eqnarray}
with
\begin{eqnarray}
f_1(x) &=& \frac{3 (x-1) \alpha(x)}{2 \pi ^2 \sqrt{(3-x) (x+1)}} \\
\frac{f_2(x)}{f_1(x)}&=& \frac{2}{x-3}-\frac{2}{x-1} + \frac{6 \,
\sqrt{x+1}}{\alpha(x) \, \sqrt{3-x}} -\frac{6 \,
\sqrt{3-x}}{\alpha(x) \,
\sqrt{x+1}} -1 \\
&&\alpha(x) = \pi -12 \arcsin \frac{\sqrt{x+1}}{2} \nonumber
\end{eqnarray}
The integrals $I_i$ can be treated by expanding the functions
$f_i=\sum_\nu f_i^{(\nu)} x^\nu$:
\begin{eqnarray}
I_i = \sum_{\nu=0}^\infty f_i^{(\nu)} c_{\text{A}}^{\nu+1}
\underbrace{\int_0^{1/c_{\text{A}}} \Exp{-z} \ z^\nu \ \D
z}_{=\nu! -\Gamma(\nu+1,1/c_{\text{A}})}.
\end{eqnarray}
All incomplete Gamma functions can be eliminated by virtue of the
recurrence relation (cf. (6.5.2) and (6.5.22) in \cite{AbramowitzStegun})
\begin{eqnarray}
\Gamma(\nu+1,1/c_{\text{A}})=\nu \Gamma(\nu,1/c_{\text{A}})+(-1)^{\nu} c_{\text{A}}^{-\nu} \Exp{-1/c_{\text{A}}} \ .
\end{eqnarray}
As is apparent from the recurrence relation, the result will be of
the form
\begin{eqnarray}
I_i = R_i(c_{\text{A}}) + \Exp{-1/c_{\text{A}}} \
S_i(c_{\text{A}}) \ . \label{FormOfI}
\end{eqnarray}
The regular part, for instance in first order of $\sB/d$,
\begin{eqnarray}
R_1(c_{\text{A}}) = c_{\text{A}} \ \frac{\sqrt{3}}{2 \pi} +
c_{\text{A}}^2 \ \frac{9-2 \sqrt{3} \pi }{3 \pi^2} +
c_{\text{A}}^3 \frac{-27+2 \sqrt{3} \pi }{3 \pi^2} + \dots \ ,
\label{FormOfIR}
\end{eqnarray}
is a series expansion about the point of jamming,
$c_{\text{A}}=0$. It is asymptotically diverging due to the
factorial which appears in the recurrence relation. Fortunately
this does not restrain us from an excellent approximation, since
for the relevant density, $\phi>\phi_{\text{o}}$, the quality of
the expansion increases for more than 10 terms in the expansion
(cf. panel (a) of the Fig.~\ref{ConvRadius}).
\\
The second part in (\ref{FormOfI}), for which the first order of
$\sB/d$ is given by
\begin{eqnarray}
S_1(c_{\text{A}}) = c_{\text{A}} \ \frac{\sqrt{3}}{2 \pi} +
c_{\text{A}}^2 \ \frac{9-2 \sqrt{3} \pi }{3 \pi^2} +
c_{\text{A}}^3 \frac{-27+2 \sqrt{3} \pi }{3 \pi^2} + \dots \ ,
\label{FormOfIS}
\end{eqnarray}
and has a positive radius of convergence (cf. panel (b) in
Fig.~\ref{ConvRadius}). This part is over-exponentially suppressed
by the prefactor $\exp{-1/c_{\text{A}}}$ close to jamming.

In the application to wet granular matter the sub-leading order $(\sB/d)^3<4 \cdot 10^{-4}$ is negligible
for a realistic value of $s \leq \scrit \approx 0.07 d$, whereby we have the concise result
\begin{eqnarray}
g_{\text{B}} (\sB) &=& {\cal N} \Exp{-z_{\text{B}}} z^2_{\text{B}} + {\cal
O}\left(z_{\text{B}}^3\right) \label{gB_Leading_Order}
\end{eqnarray}
with the abbreviation $z_{\text{B}}=\left((1+\sB
/d)^2-1\right)/c_{\text{B}}$ and $1/c_{\text{B}} \approx \phi \gc$.
The normalization $8 \phi \frac{c_{\text{B}}}{2} \int g_{\text{B}} \ \D
z_{\text{B}}=2$ according the two B-neighbors determines $\cal N$ in (\ref{gB_Leading_Order}).
Hence the result (\ref{ShortRangeCorrB}).

\section{Explicit Expressions for the Pair Correlation in Two Dimensions without free Parameters} \label{AppendixExplicitPairCorrelation}

In our general derivation of the theory of wet granular matter we
distinguished between the jamming density $\phi_{\text{J}}$ and
the (highest possible) crystalline packing
$\phi_{\text{max}}=\pi/(2\sqrt{3})$ achieved in monodisperse
domains. The exact value of the jamming density $\phi_{\text{J}}$
depends on many details such as the distribution of polydispersity
and the jamming protocol for the increase of density. When we want
to give explicit results without free parameters on the bridge
coordination $K(T,\phi,\scrit)$ and the equation of state
$P=P(T,\phi,\scrit)$ we do this for weak polydispersity, where the
difference between $\phi_{\text{J}}$ and $\phi_{\text{max}}$ is
negligible and the limiting case of 'dry' discs has been studied
extensively.

\subsection{High Density} \label{DenseContactLit}
For monodisperse 'dry' discs, $\phi_{\text{J}}=\phi_{\text{max}}$,
there are higher order corrections to the free volume result
(\ref{FreeVolumeGC}) available in the literature which are
incorporated in the final results on the bridge coordination and
the equation of state for wet granular matter. These corrections
are expansions with respect to $x=\phi_{\text{J}}-\phi$  fitted to
simulations:
\begin{eqnarray}
g^{\text{dense}}_{\text{c}} &=& \left(\frac{1}{x}+ a_0 + a_2 x^2 +
\dots \right) \ \frac{\phi_{\text{J}}} {\phi} =
\left(\frac{1}{x}+ a_0 + a_2 x^2 + \dots \right) \ \left( 1+
\frac{x}{\phi_{\text{J}}} + \dots \right)\ , \label{DensePP}
\end{eqnarray}
Equation~(\ref{DensePP}) holds in the dense regime,
$\phi_{\text{o}} < \phi < \phi_{\text{max}}$, above
$\phi_{\text{o}}=0.71$.
The numerical coefficients are $a_0=-1.07$ and $a_2=5.89$
\cite{LudingZustandsgleichung}, confirmed by our own simulations.
Similar empirical expressions are also available for polydisperse
discs in the glass state (Eq.~(6) in \cite{Donev2007}).

\subsection{Low and Moderate Density} \label{DiluteContactLit}
For the analytic treatment an explicit expression for the contact
correlation $\gc$ in Eq.~(\ref{ShortRangeCorrDepl}) is needed (as
the counterpart to the dense expression (\ref{DensePP})).  Aside
from the trivial one-dimensional case~\footnote{The configuration
space of the one-dimensional gas is $L-N {d}$, so that the
equation of state is $P (L-N d) = N T$. Comparison with the
general expression $P = \gcwall n T$ yields
$\gcwall=(1-\phi)^{-1}$.}, exact expressions for the contact
correlation of hard spheres are unknown for the dilute regime. Yet
there are well-established approximations in the literature
resulting from Scaled Particle theory \cite{SPT,SPT2}, from the
virial expansions \cite{VirialExp}, as solutions of the
Percus-Yevick closure \cite{HansenMcDonald}, as well as heuristic
expressions \cite{HeurisicContactCorrelations} such as the
Carnahan-Starling formula with corrections to better fit
simulation results (cf. Tab.~\ref{gTab}).
\begin{table*}
\centering
\begin{tabular}{c|cc|cc}\hline
  & \multicolumn{2}{c|}{Scaled Particle Theory} & \multicolumn{2}{c}{Heuristic Fits} \\[4 pt]
D & {       $\gc      $ }
  & {       $\gcwall  $ }
  & {       $\gc      $ }
  & {       $\gcwall  $ } \\[4 pt] \hline
1 & {\LARGE $\frac{1 }{1-\phi} $ }
  & {\LARGE $\frac{1 }{1-\phi} $ }
  & {\LARGE $\frac{1 }{1-\phi} $ }
  & {\LARGE $\frac{1 }{1-\phi} $ } \\[12 pt]
2
  & {\LARGE $\frac{1-\phi/2}{\left(1-\phi\right)^2}                                                  $ }
  & {\LARGE $\frac{1}{\left(1-\phi\right)^2}                                                         $ }
  & {\LARGE $\frac{1-7 \phi / 16 }{\left(1-\phi\right)^2} - \frac{\phi^3/128}{\left(1-\phi\right)^4} $ }
  & {\LARGE $\frac{1 + \phi^2 / 8}{\left(1-\phi\right)^2} - \frac{\phi^4/64} {\left(1-\phi\right)^4} $ } \\[12 pt]
3
  & {\LARGE $\frac{1- \phi / 2 + \phi^2/4}{\left(1-\phi\right)^3} $}
  & {\LARGE $\frac{1 + \phi + \phi^2}{\left(1-\phi\right)^3}      $}
  & {\LARGE $\frac{1-\phi /2 }{\left(1-\phi\right)^3}          $}
  & {\LARGE $\frac{1 + \phi + \phi^2 - \phi^3}{\left(1-\phi\right)^3}         $} \\[12 pt] \hline
\end{tabular}
\caption{The particle-particle correlation $\gc$ and the
particle-wall correlation $\gcwall$ at contact for different
spatial dimensions valid up to moderate densities. The center
column shows the results of the Scaled Particle Theory and the
right column contains the exact expression for one dimension, and
heuristic expressions \cite{HendersonReview} of Henderson
\cite{Henderson1975} for two dimensions and Carnahan-Starling
\cite{CarnahanStarling} in three dimensions.} \label{gTab}
\end{table*}

As in the dense regime \ref{DenseContactLit} we shall use the Henderson-Luding
expression \cite{LudingZustandsgleichung}
\begin{eqnarray}
g^{\text{dilute}}_{\text{c}} = \frac{1-7 \phi / 16
}{\left(1-\phi\right)^2} -
\frac{\phi^3/128}{\left(1-\phi\right)^4}  \label{DilutePP}
\end{eqnarray}
for the uncaged regime, $0 < \phi<\phi_{\text{o}}$, and the
merging function $ m(\phi) =
1/(1+\exp((\phi_{\text{o}}-\phi)/m_0)) $ with a cross-over width
$m_0=0.0111$ to smoothly connect the dense (\ref{DensePP}) and
dilute (\ref{DilutePP}) expressions
\cite{LudingZustandsgleichung}:
\begin{eqnarray}
g(s)=m(\phi) \ g^{\text{dilute}}(s) + (1-m(\phi)) \
g^{\text{dense}}(s)
\end{eqnarray}
with $g^{\text{dilute}}(0)=g^{\text{dilute}}_{\text{c}}$ and
$g^{\text{dense}}(0)=g^{\text{dense}}_{\text{c}}$ as given by the
Eqs.~(\ref{DensePP}) and (\ref{DilutePP}). The near-contact decay
has been established in the Eqs.~(\ref{ShortRangeCorr}) and
(\ref{ShortRangeCorrB}) for $\phi>\phi_{\text{o}}$, and in
Eq.~(\ref{ShortRangeCorrDepl}) for $0 < \phi<\phi_{\text{o}}$:
\begin{eqnarray}
g^{\text{dilute}}(s)&=& \gc \ \us_{\text{dilute}} \ \text{and} \\
g^{\text{dense}}(s) &=& \gc \ \us_{\text{dense}} = \gc  \ \us_{\text{dilute}}
 \left[1 +
\left(\phi \gc \ \frac{s}{d}\right)^2 \right] \label{ShortRangeGammaDense}
\end{eqnarray}
up to leading order in $\scrit$ with
\begin{eqnarray}
\us(s)_{\text{dilute}} = \exp{\left( - \phi \ \gc \
\left[\left(1+\frac{s}{d} \right)^2 -1 \right] \right)} \ . \label{ShortRangeGammaDilute}
\end{eqnarray}

With the contact expressions~(\ref{DensePP}, \ref{DilutePP}), as
well as the short-range decay formulas~(\ref{ShortRangeCorr},
\ref{ShortRangeCorrDepl}), we have sufficient information on the
dry system over the entire density range. We may therefore proceed
by introducing the hysteretic capillary bridges.


\section{Self-Consistency of Bridge Coordination $K$} \label{AppendixSelfConsistentK}
All results presented so far on the coordination $K(\phi,T)$ and
pressure $P(\phi,T)$ allowed explicit analytic results. Here we
want to demonstrate how to treat more complicated source terms of
the hysteretic system (\ref{HystereticSystem}) numerically. Such
an extension of the theory could be motivated as follows. The
current of free (unbound) approaching particles could be a
function of the free density $n_{\text{free}}$ instead of the mean
density, since some of the unconnected neighbors traverse the
voids between clusters, so that Eq.~(\ref{E3}) is changed to
\begin{eqnarray}
\phi \ \gunc + \phi \ \gunr / \us(\scrit) &=&
({1-K/K_{\text{sites}}}) \ \phi_{\text{free}} \
\gcHSS(\phi_{\text{free}}) \ . \label{E3Altered}
\end{eqnarray}
Obviously this approach is a lower estimate for the current of
freely approaching particles, which is why (\ref{E3Altered}) is
considered as a methodical example rather than a physical
competitor to the theory presented above.

\begin{figure} \begin{center}
\scalebox{0.7}{\epsffile{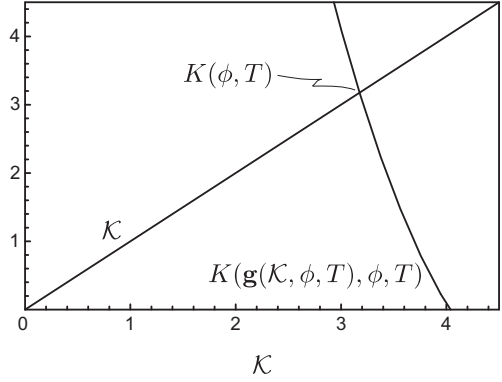}} \caption{A
typical graphical solution of the self-consistent equation
(\ref{SelfConsistentK}). Here the density is chosen to be
$\phi=0.6$ and the granular temperature is $T=0.2 \Ecb$.}
\label{chpEqStGraphic7}
\end{center} \end{figure}

With the altered Eq.~(\ref{E3Altered}) the hysteretic system
(\ref{HystereticSystem}) can still be solved analytically to find
the correlation coefficients ${\bf g}=\{ \gunc, \gbnc, \gbpc,
\gupr, \gbpr, \gunr \}$. Unlike before, due to the coupling
(\ref{E3Altered}) and the
Eqs.~(\ref{Kfrozen})-(\ref{DualFluidCond2}), the correlations
${\bf g}$ are a highly nonlinear function of $K$. Therefore
Eq.~(\ref{BridgeCoordination}) becomes a nonlinear self-consistent
equation:
\begin{eqnarray}
K({\bf g}({\cal K},\phi,T),\phi,T)={\cal K} \ .
\label{SelfConsistentK}
\end{eqnarray}
The physical value $K(\phi,T)$ of the coordination is the solution
${\cal K}$ of (\ref{SelfConsistentK}). The numerical solution of
(\ref{SelfConsistentK}) is found to be very robust, as
Fig.~\ref{chpEqStGraphic7} indicates. Plugging the resulting
self-consistent $K(\phi,T)$ back into the equation for the
pressure (\ref{FinalPressureExpression}) of wet granular matter,
we find that the critical point is shifted from
$T_{\text{c}}=0.273(5)\Ecb$ to $T_{\text{c}}=0.216(5) \Ecb$. This
reduction of the critical temperature is intuitively clear since
with less particles arriving to form bonds, the wet granular
matter 'evaporates' at lower granular temperatures.




\bibliography{ChefBibliography,MyBibliography}

\end{document}